\documentclass[]{jfm}

\usepackage{graphicx}
\usepackage{newtxtext}
\usepackage{newtxmath}
\usepackage{natbib}
\usepackage{hyperref}
\hypersetup{
    colorlinks = true,
    urlcolor   = blue,
    citecolor  = black,
}

\newcommand{\RomanNumeralCaps}[1]
\linenumbers
\usepackage{multirow}

\title{Direct numerical simulation of nucleate boiling with a resolved microlayer and conjugate heat transfer}

\author{T. Long\aff{1,2},
  J. Pan\aff{2}\corresp{\email{pan.jieyun@dalembert.upmc.fr}},
  E. Cipriano\aff{3},
  M. Bucci\aff{4},
 \and S. Zaleski\aff{2,5}}

\affiliation{\aff{1} School of Aerospace, Xi'an Jiaotong University, Xi'an, 710049, PR China
\aff{2} Institut Jean Le Rond d’Alembert UMR 7190, Sorbonne Université and CNRS, Paris, 75005, France
\aff{3}CRECK Modeling Lab, Department of Chemistry, Materials, and Chemical Engineering “G. Natta”, Politecnico di Milano, Piazza Leonardo da Vinci, 32, Milano, 20133, Italy
\aff{4}Nuclear Science and Engineering Department, Massachusetts Institute of Technology, Massachusetts, 02139, United States of America
\aff{5}Institut Universitaire de France, Paris, 75005, France
}

\begin{document}
\maketitle

\begin{abstract}
In this paper, a phase-change model based on a geometric Volume-of-Fluid (VOF) framework is extended to simulate nucleate boiling with a resolved microlayer and conjugate heat transfer. Heat conduction in both the fluid and solid domains is simultaneously solved, with Interfacial Heat-Transfer Resistance (IHTR) imposed. The present model is implemented in the open-source software Basilisk with adaptive mesh refinement (AMR), which significantly improves computational efficiency. However, the approximate projection method required for AMR introduces strong oscillations within the microlayer due to intense heat and mass transfer. This issue is addressed using a ghost fluid method, allowing nucleate boiling experiments to be successfully replicated. Compared to previous literature studies, the computational cost is reduced by three orders of magnitude. The influence of contact angle is further investigated, revealing consistent thermodynamic effects across different contact angles. Finally, a complete bubble cycle from nucleation to detachment is simulated, which, to our knowledge, has not been reported in the open literature. Reasonable agreement with experimental data is achieved, enabling key factors affecting nucleate boiling simulations in the microlayer regime to be identified, which were previously obscured by limited simulation time.
\end{abstract}

\begin{keywords}
Authors should not enter keywords on the manuscript, as these must be chosen by the author during the online submission process and will then be added during the typesetting process (see \href{https://www.cambridge.org/core/journals/journal-of-fluid-mechanics/information/list-of-keywords}{Keyword PDF} for the full list).  Other classifications will be added at the same time.
\end{keywords}


\section{Introduction}
\label{sec1:introduction}

\begin{figure}
  \centerline{\includegraphics[width=0.8\textwidth]{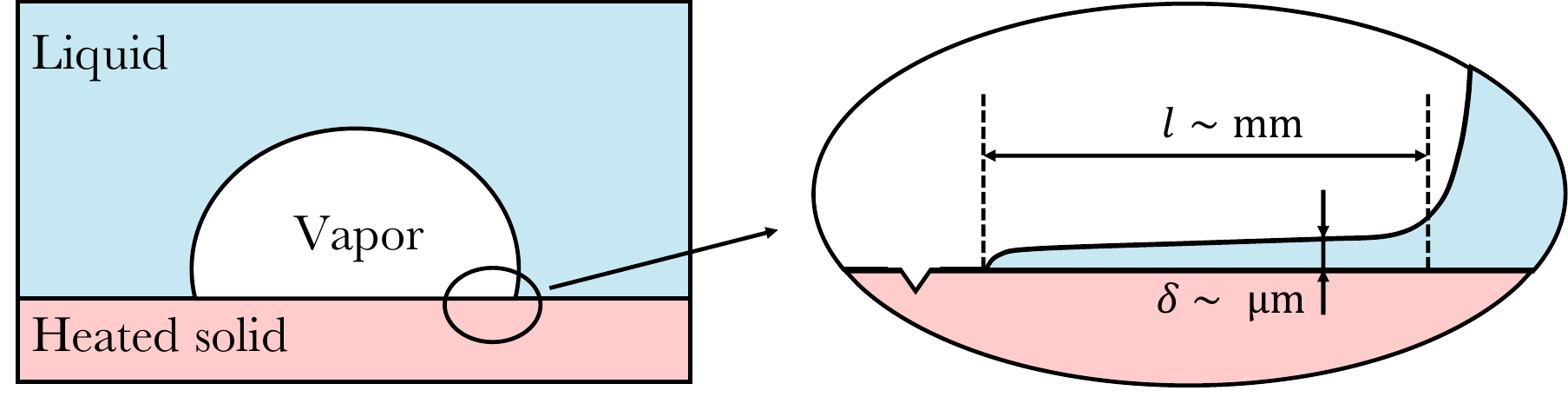}}
  \caption{Schematic of nucleate boiling with a microlayer (not to scale).}
\label{Fig_schematic_microlayer}
\end{figure}

Nucleate boiling is recognized as one of the most efficient heat-transfer processes due to the significant latent heat of evaporation \citep{chen2024review}. Consequently, it plays a critical role in various industrial applications, such as nuclear reactors \citep{manglik2006advancements}, electronics cooling \citep{cheng2017fundamental}, and thermal management subsystems in aerospace engineering \citep{dhruv2019formulation}. In nucleate boiling, a very thin liquid layer, known as the microlayer, may form between the heated wall and the liquid-vapor interface during bubble growth \citep{hansch2019microlayer}. This bubble growth regime is referred to as the microlayer regime \citep{urbano2018direct}. As depicted in figure \ref{Fig_schematic_microlayer}, the microlayer, which is merely a few microns thick, extends radially up to a few millimeters. The small thickness of the microlayer leads to very high heat flux \citep{zupanvcivc2022wall}, and its considerable radial extent over the heated surface results in a significant contribution to the overall heat transfer \citep{yabuki2014heat}. In addition, the drying of the microlayer leads to the spreading of dry spots, making its dynamics crucial for understanding Critical Heat Flux (CHF) \citep{zhao2002unified}, which is the maximum heat flux that can be transferred through nucleate boiling. These features of the microlayer have motivated intense scientific investigation into it.

In recent decades, various modern high-resolution experimental techniques have been developed, significantly advancing the understanding of the dynamics and heat-transfer characteristics of the microlayer \citep{utaka2013microlayer, jung2014experimental, bucci2016mechanistic}. Despite these advances, experimental approaches still face several limitations and shortcomings. For example, direct measurement of microlayer thickness using laser interferometry \citep{jung2018hydrodynamic, narayan2021non} inevitably introduces systematic errors due to the loss of the first fringes near the contact line, where the interface slope exceeds the observable limit \citep{tecchio2024microlayer}. In addition, the inherent uncertainties in experiments require careful evaluation \citep{kim2020assessing}. The range of controllable parameters in experiments is also limited by achievable laboratory conditions \citep{hansch2019microlayer}, which may affect the generality of experimental findings.

Compared with experimental approaches, numerical simulations are more flexible in changing operating conditions and can provide comprehensive physical information \citep{burevs2022comprehensive}. Nowadays, with rapid advances in high-performance computing, direct numerical simulation (DNS) has become a powerful tool for studying nucleate boiling and has gained increasing interest in academia \citep{chen2024review}. Despite remarkable progress in recent years, DNS of nucleate boiling in the microlayer regime remains a significant challenge due to the inherent multiscale nature of the problem. To capture the entire bubble, the domain must extend to several millimeters, while grid sizes below one micron are required to explicitly resolve the microlayer \citep{urbano2018direct}. Although a number of microlayer models have been developed to conduct multiscale simulations of nucleate boiling \citep{sato2015depletable, chen2023modeling}, their application to flow boiling, which involves bubble slipping and coalescence, is not straightforward \citep{chen2024review}. In contrast, once a DNS code is validated, it should be applicable across different scenarios. Without accounting for heat transfer, \citet{hansch2016hydrodynamics} first performed a pioneering DNS study on the hydrodynamics of microlayer formation in nucleate boiling. In their simulations, the bubble-growth rate was calculated from the analytical solution of \citet{scriven1959dynamics}. Subsequently, several researchers have studied the hydrodynamics of the microlayer using the same modeling strategy, except that the bubble-growth rate may be deduced from other models or experimental results. \citet{guion2018simulations} performed simulations over a broad range of conditions and identified the minimum set of dimensionless parameters governing the hydrodynamics of microlayer formation. \citet{giustini2020modelling} simulated microlayer formation in the boiling of industrially relevant fluids, whose properties differ significantly from typically studied fluids such as water and ethanol. Following this, \citet{giustini2024hydrodynamic} studied the formation of a dewetting ridge near the contact line during bubble growth. Most recently, \citet{saini2024direct} investigated microlayer formation during heterogeneous bubble nucleation triggered by a sudden drop in ambient pressure.

In contrast to previous studies, \citet{urbano2018direct} were the first to successfully simulate nucleate boiling in the microlayer regime with resolved thermal effects. A rigorous parametric study was carried out to determine the criterion for microlayer formation. \citet{zhang2023direct} developed an unstructured-mesh-based solver to simulate microlayer formation and evaporation driven by the local temperature gradient. This solver was later extended to study nucleate boiling on surfaces with micro-pillars \citep{zhang2023directpillar}, where the microlayer may be disturbed or disrupted by the pillars. It should be noted that in these works \citep{urbano2018direct, zhang2023direct, zhang2023directpillar}, only heat transfer in the fluid is considered, while conjugate heat transfer between the fluid and solid, which has been shown to be important in microlayer evaporation \citep{sato2015depletable, ding2018evaluation}, is neglected. \citet{hansch2019microlayer} extended their previous work \citep{hansch2016hydrodynamics} to simulate the depletion of the microlayer, where the evaporation of the microlayer is determined based on the solid temperature obtained by solving the conjugate heat transfer. However, in this work, the bubble growth rate is still approximated using the Scriven solution \citep{scriven1959dynamics} instead of solving the heat transfer in the liquid domain. Subsequently, significant progress was made by \citet{burevs2022comprehensive}, who were the first to perform detailed DNS studies of nucleate boiling in the microlayer regime by explicitly resolving heat and mass transfer in the fluid, along with conjugate heat transfer. They simulated an experiment conducted at the Massachusetts Institute of Technology (MIT) \citep{bucci2020thesis}, and their numerical results were in good agreement with experimental measurements. The same experiment was later adopted by \citet{torres2024coupling} and \citet{el2024numerical} to validate their computational models with resolved conjugate heat transfer. However, the microlayer was not the focus of \citet{torres2024coupling} and \citet{el2024numerical} and it was not studied in detail.

Although the above works demonstrated the capabilities of DNS solvers in modeling the microlayer in nucleate boiling, the computational cost of DNS still limits its application. As discussed earlier, the multiscale nature of nucleate boiling requires a large computational domain with a small grid size, leading to significant computational costs. Despite the efforts by \citet{burevs2022comprehensive} to stretch the grid and minimize the computational burden, the simulations remained highly time-consuming. For a simulation at the finest resolution, 336 cores were used, requiring approximately 400,000 CPU hours to advance the physical time to 0.5 ms \citep{burevs2022comprehensive}. By contrast, a complete bubble cycle from nucleation to detachment is typically on the order of ~10 ms. In fact, to the best of our knowledge, DNS of a complete bubble cycle with a resolved microlayer and conjugate heat transfer has not been achieved before due to the high computational cost. Employing a quad/octree-based AMR method, the free software Basilisk \citep{popinet2009accurate, popinet2015quadtree} provides a highly efficient framework for the DNS of multiphase flows \citep{pan2023edge, wang2023analysis}. Several phase-change models \citep{long2024edge, zhao2022boiling, cipriano2024multicomponent} have been developed using Basilisk, demonstrating its potential for studying complex boiling problems. In this work, we aim to extend the previous open-source phase-change model developed by \citet{cipriano2024multicomponent} to include conjugate heat transfer and to simulate nucleate boiling with a microlayer. In addition to incorporating conjugate heat transfer, the original phase-change model requires careful modifications to correctly capture the physics within the microlayer. This is due to the approximate projection method required by AMR, which leads to unphysical oscillations in phase-change problems as the velocity is discontinuous across the interface \citep{zhao2022boiling, long2024edge}. Although the previous model of \citet{cipriano2024multicomponent} appears to perform well in benchmark tests, the problem of interest in this paper is much more challenging due to the intense heat and mass transfer within the microlayer. We will demonstrate that the unphysical oscillations induced by the approximate projection method lead to incorrect microlayer dynamics, resulting in inaccurate predictions of heat and mass transfer. We adopt the ghost fluid method \citep{tanguy2014benchmarks} to address this issue, in which the singularity is removed by setting the ghost velocity according to the jump condition. By doing so, we achieve highly efficient and accurate DNS of nucleate boiling in the microlayer regime. With the present solver, we have successfully simulated the experiment conducted at MIT \citep{bucci2020thesis}. The results of the present solver not only agrees well with the numerical results of \citet{burevs2022comprehensive}, but also achieve a remarkable reduction in computational cost by three orders of magnitude, while significantly enhancing stability. The influence of the contact angle is further investigated by analyzing the contributions of different regions of the bubble to evaporation for various contact angles. Moreover, a complete bubble cycle for nucleate boiling in the microlayer regime is directly simulated with all effects explicitly resolved, which, to our knowledge, is the first such study reported in the open literature.

The remainder of this paper is structured as follows: In section \ref{sec2:numerical_method}, we briefly review the original phase-change model of \citet{cipriano2024multicomponent}. Then, we introduce the implementation of the ghost fluid method and the treatment of the solid. The extended model is verified in section \ref{sec3:verification}, followed by the numerical results and related discussions presented in section \ref{sec4:results_discussion}. Finally, concluding remarks are provided in section \ref{sec5:conclusion}.

\section{Numerical method}
\label{sec2:numerical_method}

\subsection{Governing equations}
\label{sec2.1:governing_equations}
In nucleate boiling simulations, the fluid domain is occupied by the liquid and vapor phases, which are separated by a zero-thickness interface, $\Gamma$. Both phases are assumed to be incompressible and monocomponent. With phase change considered, the governing equations for fluids are as follows:
\begin{align}
    \bnabla \bcdot \boldsymbol{u} &= S_{pc}, \label{Eq_mass_equation}\\
    \rho \left( \frac{\partial\boldsymbol{u}}{\partial t} + \left(\boldsymbol{u} \bcdot \bnabla\right)\boldsymbol{u}  \right) &= -\bnabla p + \bnabla \bcdot \left( \mu ( \bnabla \boldsymbol{u} + \bnabla \boldsymbol{u}^T) \right) + \rho \boldsymbol{g} + \boldsymbol{f}_\sigma, \label{Eq_momentum_equation}\\
    \rho C_p \left( \frac{\partial T}{\partial t} + \boldsymbol{u} \bcdot \bnabla T\right) & = \bnabla \bcdot (\lambda \bnabla T).\label{Eq_energy_equation}
\end{align}
Here $\boldsymbol{u}$, $\rho$, $\mu$, $\boldsymbol{g}$, $T$, $C_p$, and $\lambda$ represent the fluid velocity, density, dynamic viscosity, gravitational acceleration, temperature, specific heat, and thermal conductivity, respectively. The source term $S_{pc}$, introduced due to phase change, will be elaborated on later. The surface tension force is $\boldsymbol{f}_\sigma = \sigma \kappa \delta_s \boldsymbol{n}$, where $ \sigma$ is the surface tension, $\kappa$ the interface curvature, $\boldsymbol{n}$ the interface normal vector pointing from the liquid phase to the vapor phase, and $\delta_s$ the Dirac delta function concentrated at the interface. Following previous studies \citep{burevs2022comprehensive, torres2024coupling}, the surface tension $\sigma$ is assumed to be constant, implying that Marangoni convection effects are neglected. To resolve conjugate heat transfer, the following energy equation in the solid domain is considered:
\begin{equation}
\rho_s C_{p,s} \frac{\partial T_s}{\partial t} = \bnabla \bcdot (\lambda_s \bnabla T_s) + Q_h,
\label{Eq_solid_energy_equation}
\end{equation}
where the subscript $s$ denotes the prosperity of solid, and $Q_h$ is a volumetric power term due to electrical resistance heating.

\subsection{Jump conditions}
\label{sec2.2:jump_conditions}
The main challenge in the simulations of boiling flows is the accurate implementation of jump conditions at the interface. Let $H$ represent the Heaviside function \citep{tanguy2014benchmarks}, defined as $1$ in the liquid phase and $0$ in the vapor phase. Accordingly, the jump of a given fluid property $\phi$, such as density and viscosity, across the interface can be expressed as:
\begin{equation}
\phi = \phi_l H + \phi_v (1 - H),
\label{Eq_property_jump}
\end{equation}
where the subscripts $l$ and $v$ indicate the physical properties of the liquid and vapor, respectively. When phase change occurs, the mass balance at the interface gives
\begin{equation}
        \dot{m} = \rho_l (\boldsymbol{u}_l - \boldsymbol{u}_\Gamma )\bcdot \boldsymbol{n} = \rho_v (\boldsymbol{u}_v - \boldsymbol{u}_\Gamma)\bcdot \boldsymbol{n}, 
        \label{Eq_mass_flux}
\end{equation}
where $\dot{m}$ represents the mass flux. Introducing the jump operator $[\phi]_\Gamma = \phi_l - \phi_v$ in equation (\ref{Eq_mass_flux}), the velocity jump can be formulated as 
\begin{equation}
        [\boldsymbol{u}]_\Gamma = \dot{m}\left[ \frac{1}{\rho} \right]_\Gamma \boldsymbol{n}. 
        \label{Eq_velocity_jump}
\end{equation}
Moreover, the stress jump condition across the interface can be obtained by integrating the momentum equation (equation (\ref{Eq_momentum_equation})) and is usually given in the form of the pressure jump as:
\begin{equation}
   [p]_\Gamma = \sigma \kappa + 2\left[\mu \frac{\partial u_n}{\partial n} \right]_\Gamma - \dot{m}^2 \left[ \frac{1}{\rho} \right]_\Gamma,
   \label{Eq:pressure_jump}
\end{equation}
where $\frac{\partial u_n}{\partial n}$ is the normal derivative of the normal velocity component $u_n = \boldsymbol{u}\bcdot \boldsymbol{n}$. Note that the above jump conditions are also validate in the absence of phase change, i.e., $\dot{m} = 0$. 

\begin{figure}
    \centerline{\includegraphics[width=1.0\textwidth]{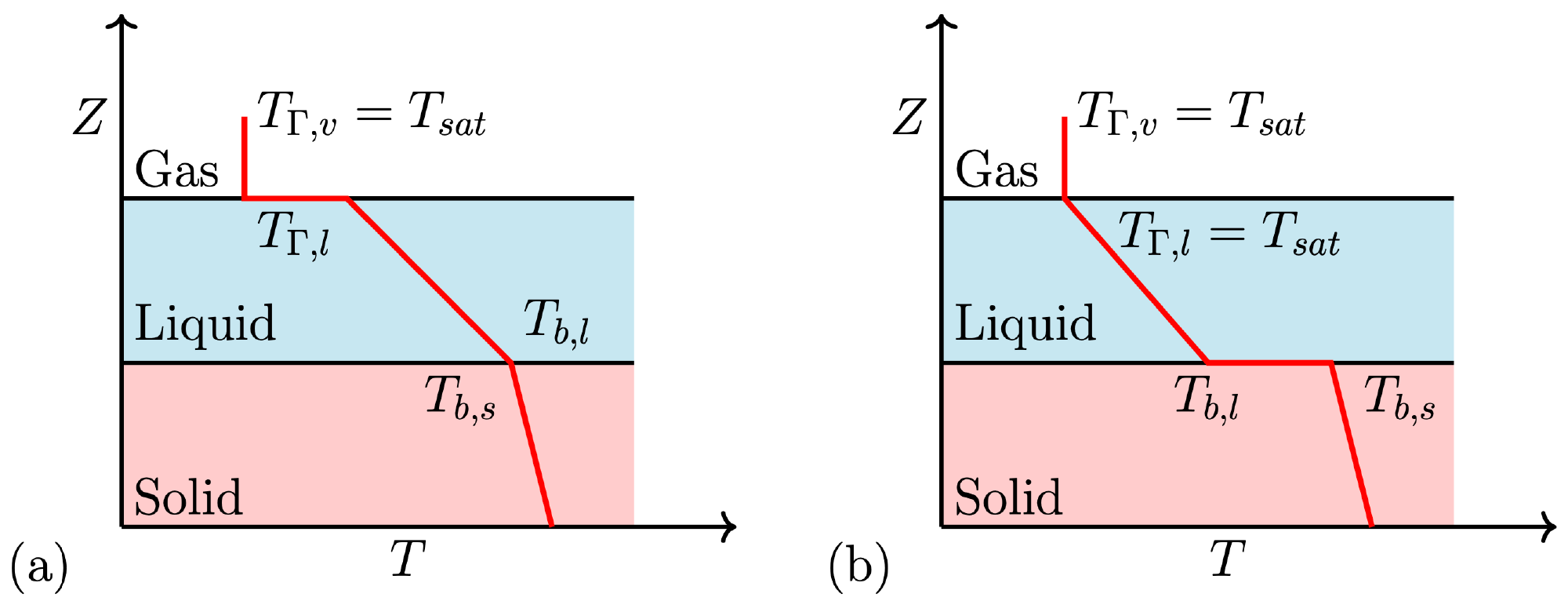}}
    \caption{Schematic for the concept of equivalent conductive resistance: (a) IHTR located at the liquid-vapor interface. (b) IHTR located at the fluid-solid boundary.}
    \label{Fig_IHTR_illustration}
\end{figure}

For the temperature field, energy conservation across the liquid-vapor interface leads to the jump condition.
\begin{equation}
    [q]_\Gamma = \dot{m}h_{lv},
    \label{Eq_jump_heatflux}
\end{equation}
where $q = \lambda \bnabla T \bcdot \boldsymbol{n}$ is the heat flux and $h_{lv}$ is the latent heat. It should be noted that thermodynamic equilibrium,
\begin{equation}
    T_{\Gamma,l} = T_{\Gamma,v} = T_{sat}, 
    \label{Eq_thermostatic_equilibrium}
\end{equation}
is assumed in the derivation of equation (\ref{Eq_jump_heatflux}) \citep{tanguy2014benchmarks}, where $T_{sat}$ denotes the saturation temperature at the ambient system pressure. However, this assumption leads to large deviations in thermodynamic characteristics from experimental observations in the presence of a microlayer \citep{giustini2016evaporative}. In fact, the degree of thermodynamic non-equilibrium within the microlayer remains an open question. Currently, a widely used modeling strategy to account for non-equilibrium effects is to introduce an additional Interfacial Heat-Transfer Resistance (IHTR) \citep{hansch2019microlayer, burevs2022comprehensive, giustini2016evaporative}. As shown in figure \ref{Fig_IHTR_illustration}(a), IHTR models assume that the vapor temperature at the liquid-vapor interface remains the saturation temperature, while the liquid temperature is higher. In the present study, a simplified IHTR model proposed by \citet{burevs2022comprehensive} is employed. Since IHTR is usually assumed to be localized in the microlayer \citep{giustini2016evaporative}, \citet{burevs2022comprehensive} demonstrated that its implementation could be simplified by introducing a numerically equivalent contact heat-transfer resistance at the fluid-solid boundary. In the presence of conjugate heat transfer, the heat flux is balanced between the solid and fluid domains \citep{torres2024coupling}, as expressed in the following equation: 
\begin{equation}
   \lambda_s \bnabla T_s \bcdot \boldsymbol{n}_s + \delta_q = \lambda_f \bnabla T_f \bcdot \boldsymbol{n}_s\quad (f=l\ \text{or}\ v),
   \label{Eqc5_heatflux_jump_solid}
\end{equation}
where $\delta_q$ denotes a Dirac source term on the solid side, and $\boldsymbol{n}_s$ is the normal vector at the fluid-solid boundary, pointing from the solid domain to the fluid domain. As shown in figure \ref{Fig_IHTR_illustration}(a), with the IHTR, the heat flux from solid to vapor is
\begin{equation}
    j_q = \frac{T_{b,s} - T_{sat}}{d/\lambda_l + R_\Gamma},
\end{equation}
where $T_{b,s}$ is the solid temperature at the fluid-solid boundary, $d$ is the microlayer thickness, and $R_\Gamma$ is the IHTR factor. Assuming the existence of a contact heat-transfer resistance $R_c$ at the fluid-solid boundary, as shown in figure \ref{Fig_IHTR_illustration}(b), the following equation is obtained:
\begin{equation}
    j_q = \frac{T_{b,s} - T_{sat}}{d/\lambda_l + R_c} = \frac{T_{b,s} - T_{sat}}{d/\lambda_l + R_\Gamma} 
\end{equation}
provided that $R_c = R_\Gamma$. In this way, the temperature discontinuity due to heat-transfer resistance is shifted from the liquid-vapor interface to the liquid-solid interface, while the overall heat flux remains unchanged. With this simplified model, only special care is needed for the implementation of conjugate heat transfer, while the original phase-change model in the fluid domain \citep{cipriano2024multicomponent}, which assumes that temperature is continuous across the interface, can be applied directly, as equation (\ref{Eq_thermostatic_equilibrium}) still holds.

\subsection{Numerical scheme}
\label{sec2.3:numerical_scheme}

\subsubsection{One-fluid method}
\label{sec2.3.1:one_fluid_method}
In the phase-change model of \citet{cipriano2024multicomponent}, a geometric VOF method with Piecewise Linear Interface Construction (PLIC) was adopted to capture the liquid-vapor interface. The VOF function $f$, defined as the volume fraction of the reference phase (which is the liquid phase in the present study) in the control cell, is updated by solving
\begin{equation}
\frac{\partial f}{\partial t}+\boldsymbol{u}_{\Gamma} \bcdot \bnabla f=0
\label{Eq_vof_advection}
\end{equation}
with a directional split advection method \citep{weymouth2010conservative, zhao2022boiling}. By considering the mass balance equation (equation (\ref{Eq_mass_flux})), the interfacial velocity is calculated as:
\begin{equation}
    \boldsymbol{u}_\Gamma = \boldsymbol{u}_l - \frac{\dot{m}}{\rho_l} \boldsymbol{n} = \boldsymbol{u}_v - \frac{\dot{m}}{\rho_v} \boldsymbol{n}.
    \label{Eq_interface_vel_onefluid}
\end{equation}
With the VOF function, the jumps of physical properties (equation (\ref{Eq_property_jump})) can be approximated by
\begin{equation}
\phi = \phi_l f + \phi_v (1 - f),
\label{Eq_average_property}
\end{equation}
which gives the so-called one-fluid method \citep{tanguy2014benchmarks}. Accordingly, the mass equation for the one-fluid velocity $\boldsymbol{u}$ can be obtained by considering the divergence-free condition in the bulk region of each phase and the velocity jump at the interface (equation (\ref{Eq_velocity_jump})), and is given by
\begin{equation}
   \bnabla \bcdot \boldsymbol{u} = \frac{\dot{m} S_\Gamma}{V_c}\left[ \frac{1}{\rho} \right]_\Gamma,
   \label{Eq_pc_source}
\end{equation}
where $V_c$ is the volume of the computational cell, and $S_{\Gamma}$ denotes the area of the interface within it.

\subsubsection{Solving the mass and momentum equations}
The phase-change model of \citet{cipriano2024multicomponent} is developed in the free software Basilisk \citep{popinet2009accurate, popinet2015quadtree}, which employs a quad/octree grid with AMR, where the velocity and pressure are collocated at cell centers. The incompressible Navier-Stokes equations are solved using a time-staggered approximate projection method, leading to the following discretization:
\begin{align}
&\rho_c^{n+\frac{1}{2}}\left(\frac{\boldsymbol{u}^*-\boldsymbol{u}^n}{\Delta t}+\left(\boldsymbol{u}^{n+\frac{1}{2}} \bcdot \bnabla\right) \boldsymbol{u}^{n+\frac{1}{2}}\right)_c=\bnabla_c \bcdot\left[\mu_f^{n+\frac{1}{2}}\left(\bnabla \boldsymbol{u}+\bnabla \boldsymbol{u}^T\right)^*\right]+\left[\left(\sigma \kappa \delta_s \boldsymbol{n}\right)^{n-\frac{1}{2}}-\bnabla p^{n-\frac{1}{2}}\right]_{f \rightarrow c}, 
\label{Eq_discretization_1}\\
&\boldsymbol{u}_c^{* *}=\boldsymbol{u}_c^*-\frac{\Delta t}{\rho_c^{n+\frac{1}{2}}}\left[\left(\sigma \kappa \delta_s \boldsymbol{n}\right)^{n-\frac{1}{2}}-\bnabla p^{n-\frac{1}{2}}\right]_{f \rightarrow c}, 
\label{Eq_discretization_2}\\
&\boldsymbol{u}_f^{*}=\boldsymbol{u}_{c \rightarrow f}^{* *} + \frac{\Delta t}{\rho_f^{n+\frac{1}{2}}}\left(\sigma \kappa \delta_s \boldsymbol{n}\right)^{n+\frac{1}{2}}, 
\label{Eq_discretization_3}\\
&\boldsymbol{u}_f^{n+1}=\boldsymbol{u}_f^{*} - \frac{\Delta t}{\rho_f^{n+\frac{1}{2}}}\bnabla p^{n+\frac{1}{2}},
\label{Eq_discretization_4}\\
&\boldsymbol{u}_c^{n+1}=\boldsymbol{u}_c^{* *}+\frac{\Delta t}{\rho_c^{n+\frac{1}{2}}}\left[\left(\sigma \kappa \delta_s \boldsymbol{n}\right)^{n+\frac{1}{2}}-\bnabla p^{n+\frac{1}{2}}\right]_{f \rightarrow c},
\label{Eq_discretization_5}
\end{align}
where the superscripts $n-\frac{1}{2}$, $n+\frac{1}{2}$, $n$, and $n+1$ represent the states at different time steps. The subscripts $c$ and $f$ denote the cell-centered and face-centered variables, respectively, and the conversion between them is achieved by a second-order accurate interpolation scheme \citep{zhao2022boiling}, denoted by the symbol $c \rightarrow f$ and $f \rightarrow c$. At each time step, the physical variables at $n-\frac{1}{2}$ and $n$ are known, and their values at $n+\frac{1}{2}$ and $n+1$ are obtained by solving the Navier-Stokes equations.

To solve the above equations, the advection term is
discretized using the Bell-Colella-Glaz (BCG) scheme \citep{bell1989second}, and the viscous
term is discretized using the implicit Crank-Nicolson scheme. Moreover, the pressure at time $n+\frac{1}{2}$ in equation (\ref{Eq_discretization_4}) is obtained by solving the Poisson equation,
\begin{equation}
\bnabla_c \bcdot\left(\frac{\Delta t}{\rho^{n+\frac{1}{2}}_f} \bnabla p^{n+\frac{1}{2}}\right)=\bnabla_c \bcdot \boldsymbol{u}_f^*-\bnabla_c \bcdot \boldsymbol{u}^{n+1}_f,
     \label{Eq_Poisson_equation}
\end{equation}
where the second term on the right-hand side is determined according to the mass equation (equation (\ref{Eq_pc_source})). This is the so-called projection step, in which the intermediate velocity $\boldsymbol{u}_f^*$ is projected onto a velocity field fulfilling the mass equation, $\boldsymbol{u}_f^{n+1}$. However, as $\boldsymbol{u}_c^{n+1}$ is interpolated using equation (\ref{Eq_discretization_5}), it is only approximately projected \citep{zhao2022boiling}. The approximate projection method is adopted to facilitate the implementation of AMR, as the cell-centered velocity is difficult to project exactly due to the spatial decoupling of the stencils used for the relaxation operator \citep{popinet2003gerris}. It has been shown that the approximate projection method introduces unphysical oscillations in the presence of phase change \citep{zhao2022boiling, long2024edge}. As shown in appendix \ref{appA:ghost_fluid_method_compare}, although the original model of \citet{cipriano2024multicomponent} performs well in many benchmark tests, it fails to correctly capture the physics of the microlayer during bubble growth. In the present work, a ghost fluid method is adopted and will be introduced later, enabling highly efficient and accurate simulations of boiling flows. 

\subsubsection{Solving the energy equation}
\begin{figure}
    \centerline{\includegraphics[width=0.8\textwidth]{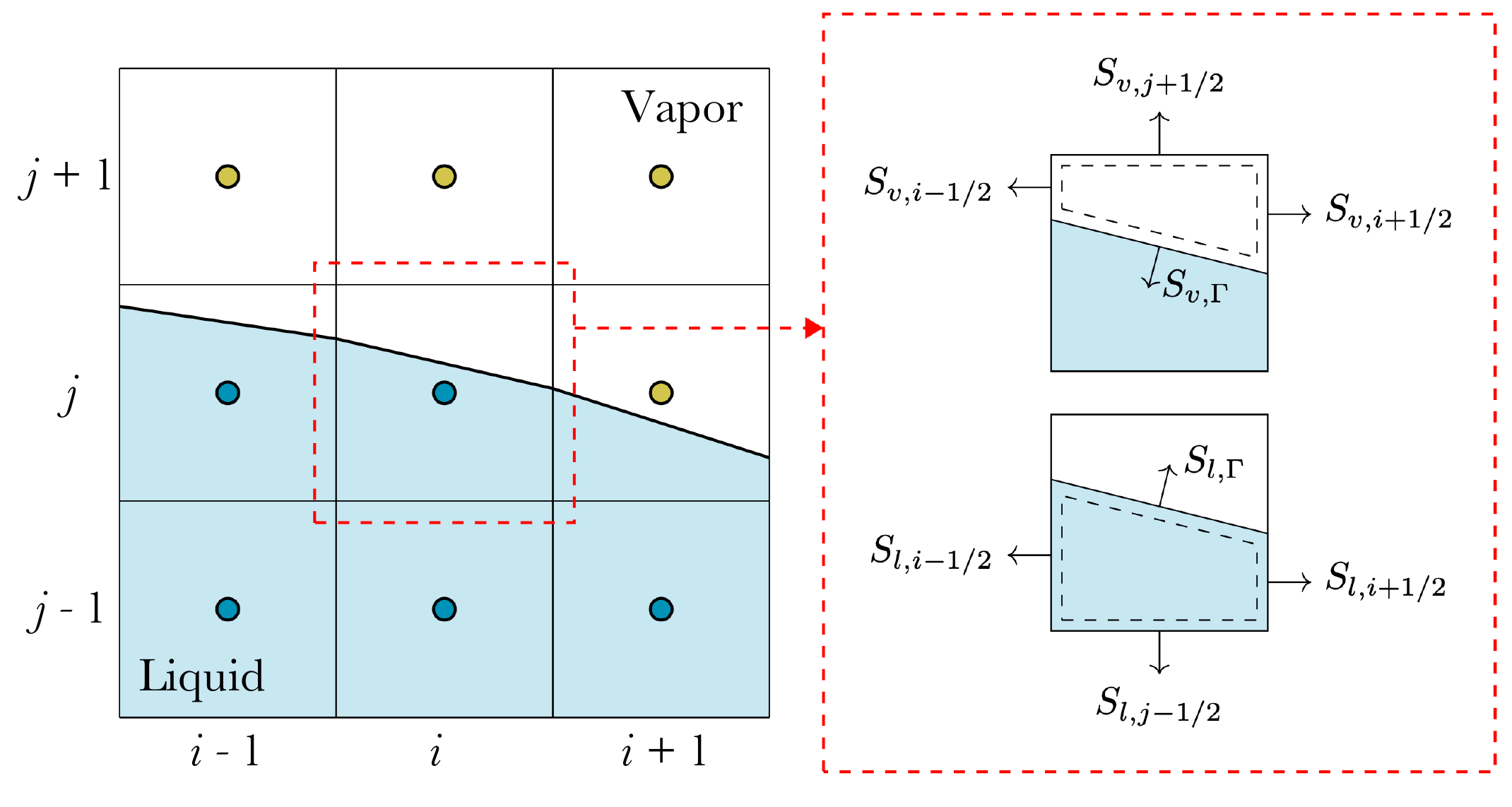}}
    \caption{Schematic for the discretization of the diffusion terms in the energy equation.}
    \label{Fig_diffusion_discretization}
\end{figure}
To solve the mass and momentum equations in the presence of phase change, the mass flux $\dot{m}$ is needed to determine the source term in equation (\ref{Eq_pc_source}). As the latent heat is usually given \textit{a priori}, the mass flux can be calculated using equation (\ref{Eq_jump_heatflux}) based on the difference between the heat fluxes across the interface. To compute the heat flux on both sides of the interface with a Dirichlet boundary condition (equation (\ref{Eq_thermostatic_equilibrium})) imposed, the normal temperature gradient $\bnabla T \bcdot \boldsymbol{n}$ is calculated using an embedded boundary method \citep{zhao2022boiling, cipriano2024multicomponent}.

To resolve the temperature evolution, the energy equation in the fluid domain is solved using the finite volume method. In the phase-change model of \citet{cipriano2024multicomponent}, two temperature fields, one for the liquid phase and one for the vapor phase, are solved separately using an operator splitting approach. This method handles the convective and diffusive parts of the energy equation independently \citep{cipriano2024multicomponent}. For brevity, the volume fractions of the liquid and vapor phases are defined as $\theta_l = f$ and $\theta_v = 1 - f$. First, the advection term is solved as:
\begin{equation}
     \left(\frac{(\theta_k T_k)^* - (\theta_k T_k)^{n-\frac{1}{2}}}{\Delta t}\right)_c = -\frac{1}{V_c} \sum_{cell\ faces}T_{k,f}^nF_{\theta_k,f}, \\
    \label{Eq_energy_advection}
\end{equation}
where $k = l\ \text{or}\ v$ indicates the phase of interest, $V_c$ denotes the cell volume, $T^n_{k,f}$ represents the face-centered temperature obtained using the BCG scheme \citep{bell1989second}, and $F_{\theta_k,f}$ is the volume flux computed during the geometrical advection of the VOF function \citep{cipriano2024multicomponent}. In this way, the energy is advected as a tracer associated with the VOF advection, avoiding numerical diffusion across the interface. 

Once we obtain the intermediate temperature field, which takes into account the advection effect, we can solve the diffusion term using finite volume discretization:
\begin{equation}
    \begin{aligned}
     \left(\rho_k C_{p,k}\theta_k^{n+\frac{1}{2}}\frac{T_k^{n+\frac{1}{2}} - T_k^*}{\Delta t}\right)_c = -\frac{1}{V_c} \left[\sum_{cell\ faces}  \lambda_k \bnabla T_{k,f}^{n+\frac{1}{2}}\bcdot \boldsymbol{S}_{k,f}  + \left(\lambda_k \frac{\partial T}{\partial n }\bigg|_{\Gamma,k} \boldsymbol{n} \bcdot \boldsymbol{S}_{\Gamma,k} \right)^{n - \frac{1}{2}}\right], \\
    \end{aligned}
    \label{Eq_energy_diffusion}
\end{equation}
where $\boldsymbol{S}_{k,f}$ and $\boldsymbol{S}_{\Gamma,k}$ denote the area vectors of the cell face and the interface segment, respectively, as illustrated in figure \ref{Fig_diffusion_discretization}. The face-centered temperature gradient $\bnabla T_{k,f}$ is computed using a second-order accurate central difference method, and the normal temperature gradient at the interface $\frac{\partial T}{\partial n }|_{\Gamma,k}$ is calculated using the embedded boundary method \citep{zhao2022boiling, cipriano2024multicomponent}. Note that the second term on the right-hand side of equation (\ref{Eq_energy_diffusion}) is introduced only in the cells cut by the interface.

\subsubsection{Ghost fluid method}
In previous sections, the original phase-change model of \citet{cipriano2024multicomponent} is briefly reviewed. In this section, the modifications and extensions for simulating nucleate boiling with a resolved microlayer and conjugate heat transfer are introduced. As mentioned earlier, the source term in the mass equation induced by phase change (equation (\ref{Eq_pc_source})) is singular and nonzero only in the interfacial cells. As a result, numerical oscillations may arise near the interface, especially for the cell-centered velocity, which is only approximately projected \citep{zhao2022boiling, long2024edge}. However, the approximate projection method is required for the implementation of AMR, which can significantly improve computational efficiency. To achieve a highly efficient and accurate phase-change model, the original model of \citet{cipriano2024multicomponent} is improved by adopting the ghost fluid method \citep{tanguy2014benchmarks}. As shown in appendix \ref{appA:ghost_fluid_method_compare}, this modification is crucial for obtaining an accurate prediction of heat transfer within the microlayer. The principle of the ghost fluid method is to solve two separate velocity fields for each phase. Let $C$ denote the color function, which is computed by 
\begin{equation}
     C = \left\{ 
     \begin{aligned}
        &1 \quad \text{if $f \geq 0.5$}, \\
        &0 \quad \text{if $f < 0.5$}. \\  
     \end{aligned}
     \right.
     \label{Eq:color_function}
\end{equation}
Accordingly, the velocity fields can be expressed as
\begin{equation}
    \begin{aligned}
     \boldsymbol{u}_l & = C\boldsymbol{u}_l + (1 - C) \boldsymbol{u}_l^g, \\
     \boldsymbol{u}_v & = (1 - C)\boldsymbol{u}_v + C \boldsymbol{u}_v^g, \\
    \end{aligned}
    \label{Eq:two_velocity}
\end{equation}
where the ghost velocities $\boldsymbol{u}_l^g$ and $\boldsymbol{u}_v^g$ are calculated as
\begin{equation}
    \begin{aligned}
     \boldsymbol{u}_{v}^{g} &= \boldsymbol{u}_{l} - \dot{m}\left[ \frac{1}{\rho} \right]_\Gamma \boldsymbol{n}, \\
    \boldsymbol{u}_{l}^{g} &= \boldsymbol{u}_{v} + \dot{m}\left[ \frac{1}{\rho} \right]_\Gamma \boldsymbol{n}. \\ 
    \end{aligned}
    \label{Eq_ghost_velocity}
\end{equation}
Since the above equation is obtained according to the velocity jump condition (equation (\ref{Eq_velocity_jump})), the singularity at the interface is removed \citep{tanguy2014benchmarks}, leading to the new mass equations:
\begin{equation}
    \bnabla \bcdot \boldsymbol{u}_l = \bnabla \bcdot \boldsymbol{u}_v = 0.
    \label{Eq_divergence-free}
\end{equation}
Consequently, during the projection step, equation (\ref{Eq_Poisson_equation}) becomes
\begin{equation}
     \bnabla_c \bcdot \left(\frac{\Delta t}{\rho^{n+\frac{1}{2}}_f}\bnabla p^{n+1} \right) = \left\{ 
     \begin{aligned}
        &\bnabla_c \bcdot \boldsymbol{u}_{f,l}^* \quad \text{if $C = 1$}, \\
        &\bnabla_c \bcdot \boldsymbol{u}_{f,v}^* \quad \text{if $C = 0$}, \\  
     \end{aligned}
     \right.
     \label{Eq_poisson_ghost_fluid}
\end{equation}
where the singular source term due to phase change on the right-hand side is eliminated. In practice, there is no need to populate ghost velocities throughout the domain. Equation (\ref{Eq_ghost_velocity}) is applied only within a narrow band near the interface, defined by whether a cell or any of its neighbors is cut by the interface.

\subsubsection{Treatment of solid}
\begin{figure}
     \centerline{\includegraphics[width=0.5\textwidth]{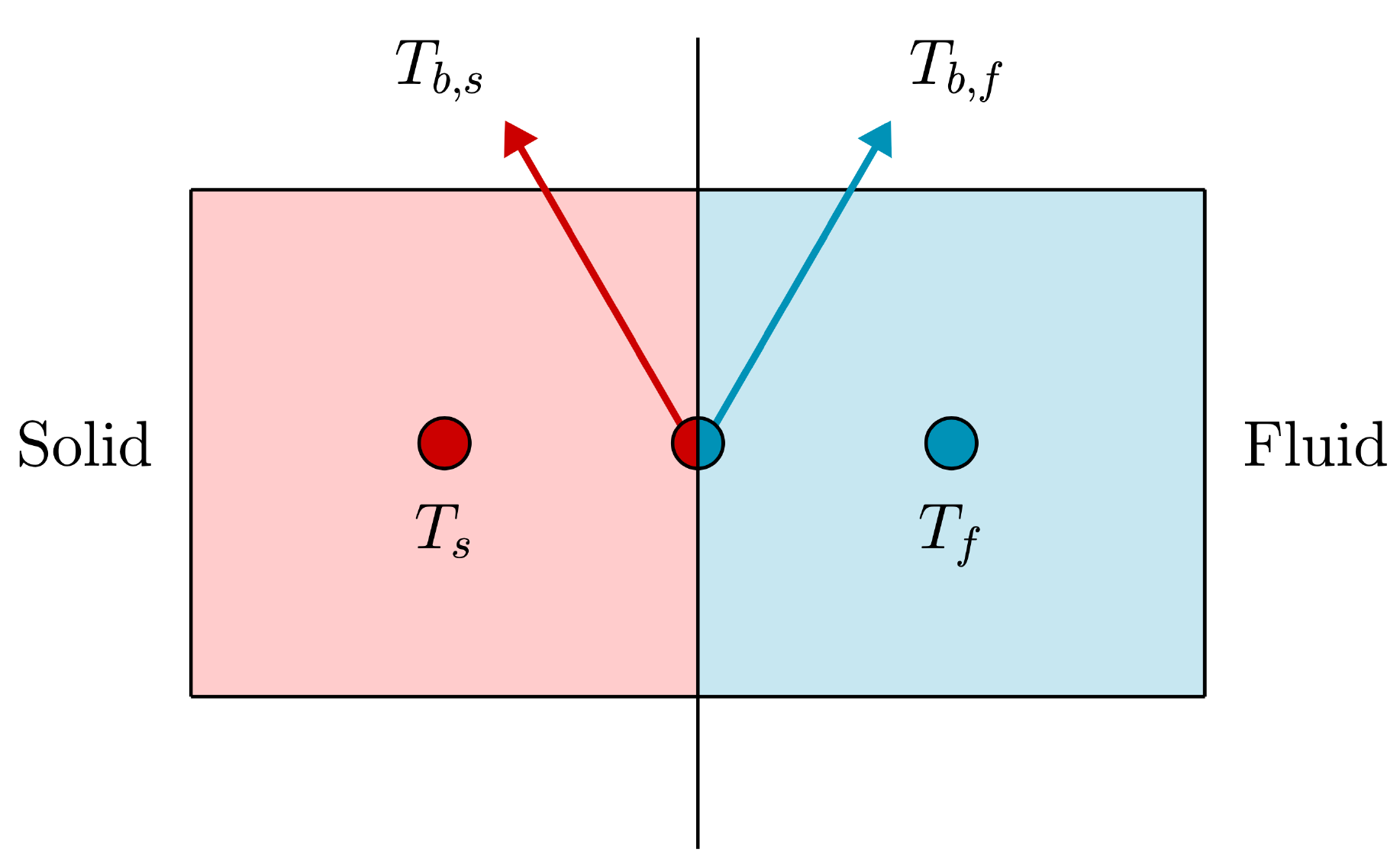}}
    \caption{Schematic for the implicit discretization of the heat diffusion terms at the fluid-solid boundary.}
    \label{Fig_diffusion_bc}
\end{figure}
Finally, the solution of heat conduction in the solid domain and the implementation of conjugate heat transfer between the fluid and solid domains are elaborated. Applying the finite volume discretization to equation (\ref{Eq_solid_energy_equation}) leads to
\begin{equation}
    \begin{aligned}
     \left(\rho_s C_{p,s}\frac{T_s^{n+\frac{1}{2}} - T_s^{n-\frac{1}{2}}}{\Delta t}\right)_c = -\frac{1}{V_c} \left(\sum_{cell\ faces}  \lambda_s \bnabla T_{s,f}^{n+\frac{1}{2}}\bcdot \boldsymbol{S}_{s,f} \right) + Q_h^{n-\frac{1}{2}}. \\
    \end{aligned}
    \label{Eq_solid_energy_diffusion}
\end{equation}
It can be seen from equation (\ref{Eq_energy_diffusion}) and equation (\ref{Eq_solid_energy_diffusion}) that the heat conduction terms are discretized using an implicit scheme in both the fluid and solid domains, eliminating the strict timestep constraint imposed by these terms. During simulations, these two equations are solved simultaneously with the associated boundary conditions. For fluid and solid cells near the boundary, the discretization stencil is incomplete when the face-centered temperature gradient is computed with a central scheme. As shown in figure \ref{Fig_diffusion_bc}, this issue is addressed by employing a one-sided difference scheme, relying on two associated boundary temperatures, $T_{b,s}$ and $T_{b,f}$. With the contact heat-transfer resistance and the continuity of heat flux, the following equation is obtained:
\begin{equation}
    2\lambda_s \frac{T_s - T_{b,s}}{\Delta} = \frac{T_{b,s} - T_{b,f}}{R_c}=  2\lambda_f \frac{T_{b,f} - T_{f}}{\Delta},
    \label{Eq_discretized_jump_conjugate}
\end{equation}
where $\Delta$ is the grid size. This equation can be simplified by introducing $R_s = \Delta / (2 \lambda_s)$ and $R_f = \Delta / (2 \lambda_f)$. The solution of equation (\ref{Eq_discretized_jump_conjugate}) yields the boundary values as follows:
\begin{equation}
\begin{aligned}
    T_{b,s} = & \frac{(R_c + R_f)T_s + R_s T_f}{R_c + R_f + R_s},\\
    T_{b,f} = &\frac{(R_c + R_s)T_f + R_f T_s }{R_c + R_f + R_s}.
\end{aligned}
\label{Eq_boundary_temperature}
\end{equation}
The temperature gradient at the fluid-solid boundary can then be calculated using equation (\ref{Eq_discretized_jump_conjugate}). Note that the continuity of temperature \citep{huber2017direct} across the fluid-solid boundary is automatically recovered when $R_c = 0$. In addition, all the above procedures can be directly applied when the solid consists of two different materials. Following \citet{burevs2022comprehensive}, the properties of the solid are mixed based on the solid volume fractions. In this study, binary mixed solids are considered, and the cell-specific values of $\lambda_s$ and $C_{p,s}$ are computed by
\begin{equation}
    \begin{aligned}
    \lambda_s &= \lambda_{1}f_{s,1} + \lambda_{2}(1 - f_{s,1}),\\
     C_{p,s} &= C_{p,1}f_{s,1} + C_{p,2}(1 - f_{s,1}),\\
    \end{aligned}
\end{equation}
where $f_{s,1}$ represent the volume fraction of material 1.

\subsubsection{Summary}
For clarity, the numerical procedures for each time step are summarized as below:
\begin{enumerate}
\item Solve the advection of the VOF function using the split scheme (equation (\ref{Eq_vof_advection})).
\item Solve the advection term of the energy equation in the fluid domain (equation (\ref{Eq_energy_advection})).
\item Solve heat conduction in the fluid and solid domains simultaneously using an implicitly coupled approach (equations (\ref{Eq_energy_diffusion}) and 
 (\ref{Eq_solid_energy_diffusion})).
\item Compute the mass flux (equation (\ref{Eq_jump_heatflux})) and set the ghost velocities (equation (\ref{Eq_ghost_velocity})).
\item Solve the mass and momentum equations (equations (\ref{Eq_discretization_1})–(\ref{Eq_discretization_5})).
\end{enumerate}

\section{Verification}
\label{sec3:verification}
\begin{figure}
    \centerline{\includegraphics[width=0.8\textwidth]{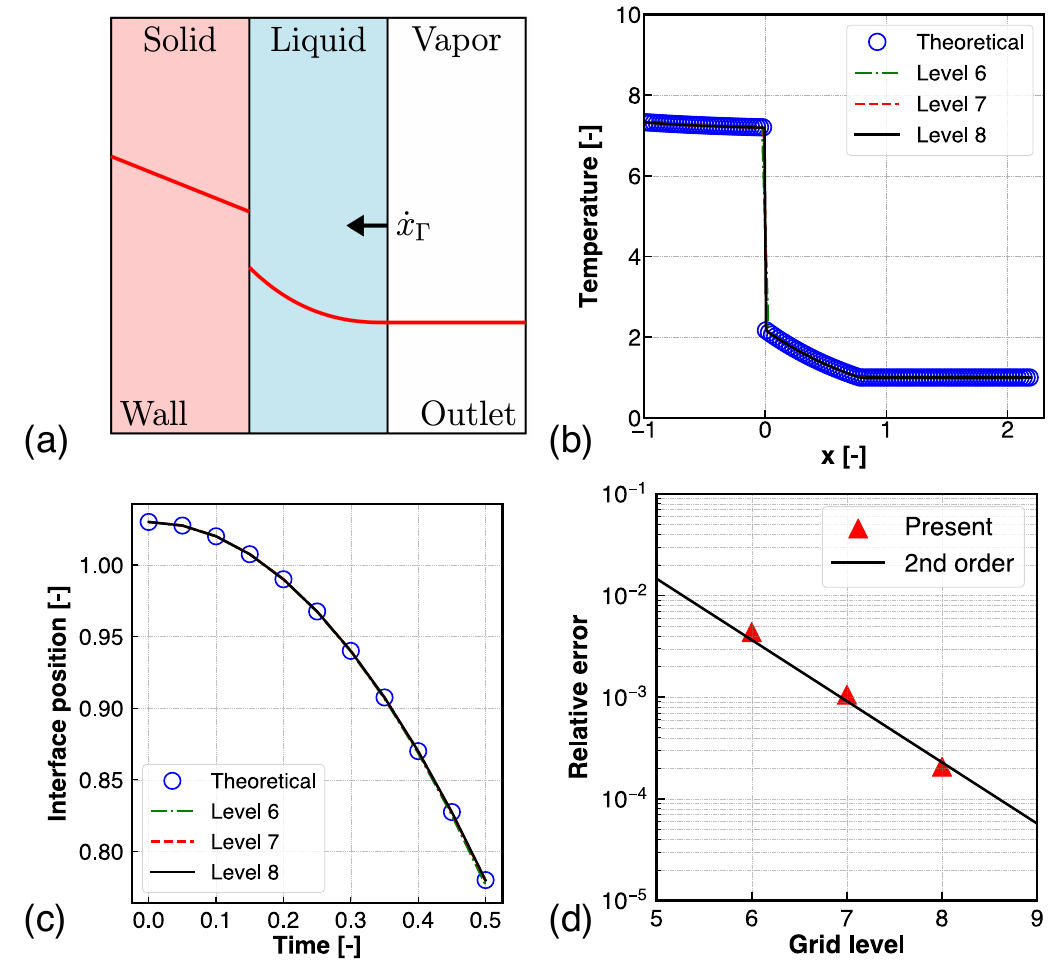}}
    \caption{Film evaporation with conjugate heat transfer: (a) Schematic of the 1D film evaporation with conjugate heat transfer. (b) Temperature distribution at \( t = 0.5 \) s. (c) Time history of the interface position. (d) Relative error of the interface position on different grid resolutions. Grid levels 6 to 8 correspond to effective grid resolutions ranging from \( 64 \times 1 \) cells to \( 256 \times 1 \) cells, resulting in minimum grid sizes from $0.05$ to $0.0125$.}
    \label{Fig_conjugate}
\end{figure}
The phase-change model of \citet{cipriano2024multicomponent} has been verified for the liquid-vapor system using a wide range of benchmark tests, with all codes available on the Basilisk website \citep{edosandbox}. Here, we focus primarily on the validation of the conjugate heat transfer extension. In this section, we consider a film evaporating in the presence of conjugate heat transfer, as illustrated in figure \ref{Fig_conjugate}(a). This benchmark case, proposed by \citet{bures2021thesis}, is specifically designed for code verification. The theoretical solution can be derived using the Method of Manufactured Solutions (MMS) \citep{roache1998verification}. Before detailing the case setup, some notations are introduced for clarity. In the following derivations, $\phi'(x,t)$ and $\dot{\phi}(x,t)$ denote the derivatives of the variable $\phi$ with respect to $x$ and $t$, respectively. Additional primes and dots indicate corresponding higher-order derivatives.

In this case, the interface position is specified as 
\begin{equation}
    x_{\Gamma}(t) = x_0 - Z t^2,
    \label{EqC5_conjugate_interface_pos}
\end{equation} 
where $x_0$ is the initial position, and $Z$ is a user-defined parameter. The vapor temperature is maintained at the saturation temperature $T_{sat}$ throughout the process, while the liquid temperature is modeled with an exponential profile:
\begin{equation}
   T_l(x, t)=T_{sat} E(x, t)=T_{sat} \exp \left[M \dot{x}_\Gamma(t)\left(x-x_\Gamma(t)\right)\right].
   \label{EqC5_conjugate_Tl}
\end{equation}
Here the factor $M$ is defined as 
\begin{equation}
   M = \frac{h_{lv} \rho_l}{T_{sat}\lambda_l},
   \label{EqC5_conjugate_M}
\end{equation}
ensuring the Stefan condition
\begin{equation}
   -\frac{\lambda_l }{h_{lv} \rho_l}T_l(x_\Gamma, t) = -\dot{x}_\Gamma(t) 
   \label{EqC5_conjugate_Stefan}
\end{equation}
is satisfied. To enforce the flux and temperature jump conditions at the fluid-solid boundary, the solid temperature distribution is given by 
\begin{equation}
T_{s}(x, t)=T_s\left[\left(\frac{\lambda_l}{\lambda_{s}}+D(t)\right) E(x, t)+C(t) E(0, t)\right],
\label{EqC5_conjugate_Ts}
\end{equation}
where $D(t)$ and $C(t)$ are defined as
\begin{equation}
D(t)=\frac{\delta_q(t)}{\lambda_{s} T_s} \frac{1}{E^{\prime}(0, t)}
\label{EqC5_conjugate_D}
\end{equation}
and
\begin{equation}
C(t)=-\left[\lambda_l R_{c} M \dot{x}_\Gamma(t)+D(t)+\left(\frac{\lambda_l}{\lambda_{s}}-1\right)\right],
\label{EqC5_conjugate_C}
\end{equation}
respectively. Here the wall source term $\delta_q(t)$ is chosen as $\delta_q(t)=-\psi \dot{x}_\Gamma(t)$, with $\psi$ being a given control parameter. To achieve the temperature evolution described above, additional source terms must be imposed in both the liquid and solid domains during the simulation, which are given by
\begin{align}
S_l(x, t) = & C_{p, l} \dot{T}_l(x, t)-\lambda_l T_l^{\prime \prime}(x, t)=T_s\left[C_{p, l} \dot{E}(x, t)-\lambda_l E^{\prime \prime}(x, t)\right], \\
S_{s}(x, t) = &C_{p,s} \dot{T}_s(x, t)-\lambda_{s} T_s^{\prime \prime}(x, t) 
\nonumber\\
= & T_s\left\{C_{p,s}\left[\left(\frac{\lambda_l}{\lambda_{s}}+D(t)\right) \dot{E}(x, t)+\dot{D}(t) E(x, t)+\dot{C}(t) E(0, t)+C(t) \dot{E}(0, t)\right]
\right. \nonumber\\
&\left.-\lambda_{s}\left(\frac{\lambda_l}{\lambda_{s}}+D(t)\right) E^{\prime \prime}(x, t)\right\} .
\end{align}

In the simulation, the fluid-solid boundary is placed at the origin, with the lengths of the fluid and solid domains chosen as $2.2$ and $1$, respectively. Following \citet{bures2021thesis}, with viscosity neglected, the physical properties are set as:
\begin{equation}
\left\{\begin{aligned}
\rho_{l}&=1, \lambda_{l}=1,C_{p,l} =2, \rho_{v}=1, \lambda_{v}=1,C_{p,v} =2,\\
\rho_{s}&=4, \lambda_{s}=7,C_{p,s} =5, T_{s}=1, h_{lv}=1, R_c = 2.3,\\
x_0&=1.03, Z=1, \psi = 1.65.\\
\end{aligned}\right.
\end{equation}
Note that all parameters above are dimensionless. The simulation is performed up to $t = 0.5$ with three grid levels ranging from $6$ to $8$, resulting in grid sizes from $0.05$ to $0.0125$. Note that for a given grid level $L$, the corresponding number of cells is $N = 2^{L}$. In figure \ref{Fig_conjugate}(b), the temperature distributions at the final time for different grid refinements are compared with the theoretical solution. It can be observed that the temperature discontinuity at the fluid-solid boundary is accurately and sharply captured by the present method. Moreover, as shown in figure \ref{Fig_conjugate}(c), the time evolutions of the interface positions for all grid levels are in good agreement with the theoretical solution. For a quantitative comparison, the relative error of the final interface position is computed by 
\begin{equation}
    E(x_\Gamma(0.5)) = \frac{|x_\Gamma(0.5) - x_\Gamma(0.5)^{num}|}{x_\Gamma(0.5)},
\end{equation}
where the superscript $num$ represents the numerical solution. As can be observed from figure \ref{Fig_conjugate}(d), the present method exhibits a second-order convergence rate, demonstrating its efficacy for phase change problems involving conjugate heat transfer.

\section{Results and discussion}
\label{sec4:results_discussion}

After validating the phase-change model with conjugate heat transfer, it is applied to simulate the experiments of \citet{bucci2020thesis}. In these experiments, the bubble growth in pool boiling of water under atmospheric conditions was investigated. Note that all codes used in this study are available on the Basilisk website \citep{tiansandbox}.

\subsection{Simulation setup}
\label{sec4.1:simulation_setup}
\begin{figure}
    \centerline{\includegraphics[width=0.5\textwidth]{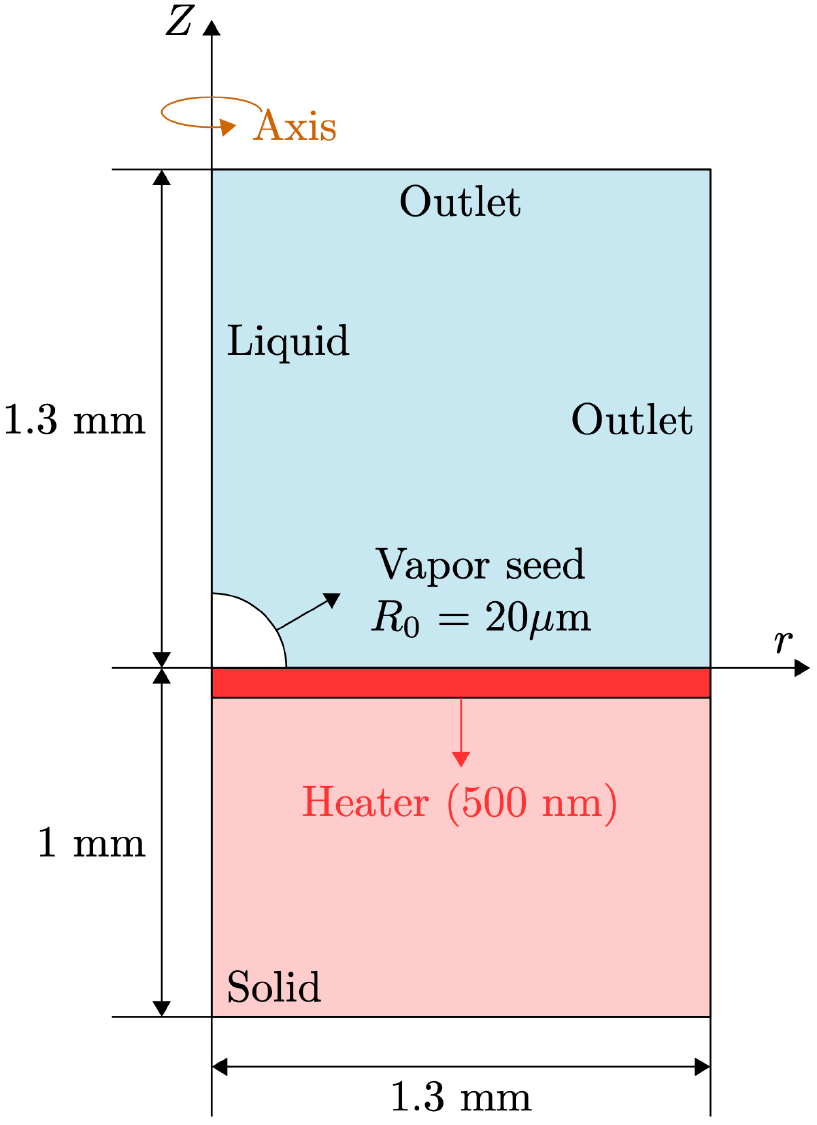}}
    \caption{Schematic of the computational domain used for the simulations of the experiment of \citet{bucci2020thesis} (not to scale). }
    \label{Fig_sketch_domain}
\end{figure}
\begin{table}
\begin{center}
\def~{\hphantom{0}}
\begin{tabular}{cccccc}
Property & Liquid & Vapor & Two-phase & Sapphire & Titanium\\
\hline
$\rho\ \left[\mathrm{kg/m^3}\right]$ & 958 & 0.598 & - & 3980 & 4510 \\
$\mu\ [\mathrm{Pa\cdot s}]$ & $2.82\times10^{-4}$ & $1.22\times10^{-5}$ & - & - & -\\
$C_{p}\ [\mathrm{J / (kg\cdot K)}]$ & 4220 & 2080 & - &929&544\\
$\lambda\ [\mathrm{W / (m\cdot K)}]$ & 0.677 & 0.0246 & - & 25.1 & 17.0\\
$h_{lv}\ [\mathrm{J/kg}]$ & - & - & $2.26\times10^{6}$ & - & - \\
$T_{sat}\ [\mathrm{K}]$ & - & - & $373.15$ & - & - \\
$\sigma\ [\mathrm{N / m}]$ & - & - & $0.0589$ & - & - \\
\end{tabular}
\caption{Physical properties used in the simulations of the experiment of \citet{bucci2020thesis}.}
\label{Tab_physical_property}
\end{center}
\end{table}
\begin{table}
\begin{center}
\def~{\hphantom{0}}
\begin{tabular}{c c c c c c c}
\multirow[t]{1}{*}{} & \multicolumn{3}{c}{ \citet{burevs2022comprehensive}} & \multicolumn{3}{c}{ Present }\\
& Coarse & Medium & Fine & Level 10 & Level 11 & Level 12 \\
\hline Minimum grid size $[\rm{\mu m}]$& 1.05 & 0.749 & 0.475 & 2.246 & 1.123 & 0.562 \\
Number of cores $[-]$& 70 & - & 336 & 4 & 24 & 48 \\
CPU time [core-h]& $\sim20000$ & - & $\sim400000$ & $\sim5$ & $\sim80$ & $\sim680$ \\
\end{tabular}
\caption{Comparison of computational efficiency between the present work and the study of \citet{burevs2022comprehensive}.}
\label{Tab_efficiency_comparison}
\end{center}
\end{table}

As mentioned above, the IHTR is considered in the simulations by imposing a temperature contact discontinuity at the fluid-solid interface. The contact heat-transfer resistance factor $R_c$ is set equal to the IHTR factor $R_\Gamma$ to preserve the overall heat flux. In the work of \citet{burevs2022comprehensive}, the Hertz–Knudsen relation is combined with the Clausius–Clapeyron relation to model $R_\Gamma$, leading to the following expression:
\begin{equation}
    R_\Gamma = \frac{1}{\omega}\frac{1}{\rho_v h_{lv}^2}\sqrt{\frac{2\pi R_g T_{sat}^3}{M_v}},
    \label{Eq_accomodation_coef}
\end{equation}
where $M_v$ is the molar mass, $R_g$ is the universal gas constant, and $\omega$ is the so-called accommodation coefficient, which is \textit{a priori} unknown. By fitting experimental data, \citet{burevs2022comprehensive} evaluated the bounds of $\omega$ and selected two values for their simulations: $\omega = 0.0345$ and $\omega = 0.0460$. In the present work, the same two accommodation coefficients are adopted, referring to the cases with $\omega = 0.0345$ and $\omega = 0.0460$ as case A and case B, respectively.

In addition to accommodation coefficients, the other physical properties considered in the simulations are listed in table \ref{Tab_physical_property}. The simulations are performed with an axisymmetric configuration, which accurately represents the real experimental conditions, where perfect axial symmetry of the growing bubble has been observed \citep{bucci2020thesis}. As shown in figure \ref{Fig_sketch_domain}, a rectangular domain of $\left[ 0\ \rm{mm}, 1.3 \ \rm{mm} \right] \times \left[ -1.0\ \rm{mm}, 1.3 \ \rm{mm} \right]$ is employed, with the fluid-solid boundary placed at $z = 0\ \mathrm{mm}$. The solid phase consists of a $1\ \rm{mm}$ thick sapphire substrate with a $500\ \rm{nm}$ thick titanium heater. The heater is modeled using a volumetric source term ($Q_h$ in equation (\ref{Eq_solid_energy_equation})), and, following \citet{burevs2022comprehensive}, we consider the case with an applied heat flux of $425\ \mathrm{kW/m^2}$. With the left boundary as the axis of symmetry, outflow boundary conditions are applied to the right and top boundaries of the fluid domain, while the fluid-solid boundary is treated as a no-slip wall. Note that since the VOF function is advected using the discretized velocity located half a grid spacing above the wall, an implicit slip condition is introduced for the interface motion \citep{afkhami2008height}. In the solid domain, a Neumann boundary condition (zero heat flux) is imposed on the right and bottom boundaries. At the fluid-solid boundary, a three-phase contact line forms, requiring the specification of a contact angle $\theta_C$. In the work of \citet{burevs2022comprehensive}, a dynamic contact angle model is adopted, yielding an angle of less than $1^\circ$ throughout the simulations. However, as will be shown in this paper, the influence of the contact angle is minimal when it is less than $10^\circ$. In this section, following \citet{el2024numerical}, a static contact angle model with $\theta_C = 5^\circ$ is used. The static contact angle is imposed using the height function approach of \citet{afkhami2008height}, which is already implemented in Basilisk.

With the setup described above, simulations have been conducted at increasing grid levels from $10$ to $12$, with corresponding minimum grid sizes ranging from $2.246\ \rm{\mu m}$ to $0.562\ \rm{\mu m}$. The number of CPU cores and the total CPU time are compared with those reported by \citet{burevs2022comprehensive} in table \ref{Tab_efficiency_comparison}. It is observed that the number of CPU cores and CPU hours required for the simulations are significantly reduced using the present model. For the finest resolution, only around 680 core-hours are needed with 48 cores, compared to approximately $400\ 000$ core-hours with 336 cores reported by \citet{burevs2022comprehensive}. The significant improvement in computational efficiency is attributed to two factors. First, for the same effective resolution, the use of AMR greatly reduces the total number of grid cells in the simulation. Second, the present model is more robust and stable, allowing for larger timestep. During the simulation, as required by numerical stability \citep{burevs2022comprehensive}, the timestep $\Delta t$ is determined by
\begin{equation}
    \Delta t = \mathrm{min}\left(C_{adv}\frac{\Delta}{u_{max}}, C_\sigma \sqrt{\frac{(\rho_v + \rho_l)\Delta ^ 3}{\sigma}}\right),
\end{equation}
where $\Delta$ is the minimum grid size used, and $u_{max}$ is the maximum velocity component in all directions. Here $C_{adv}$ and $C_\sigma$ are the limiting coefficients for the advection and capillary terms, which are set to $C_{adv} = 0.5$ and $C_\sigma = 0.282$ in the present study. In the work of \citet{burevs2022comprehensive}, the time step limits are much more stringent ($C_{adv} = 0.02$ and $C_{\sigma} = 0.063$) due to the oscillations induced by the strong heat transfer involved in this problem.

\subsection{Initial condition}
\label{sec4.2:initial_condition}
\begin{figure}
     \centerline{\includegraphics[width=0.9\textwidth]{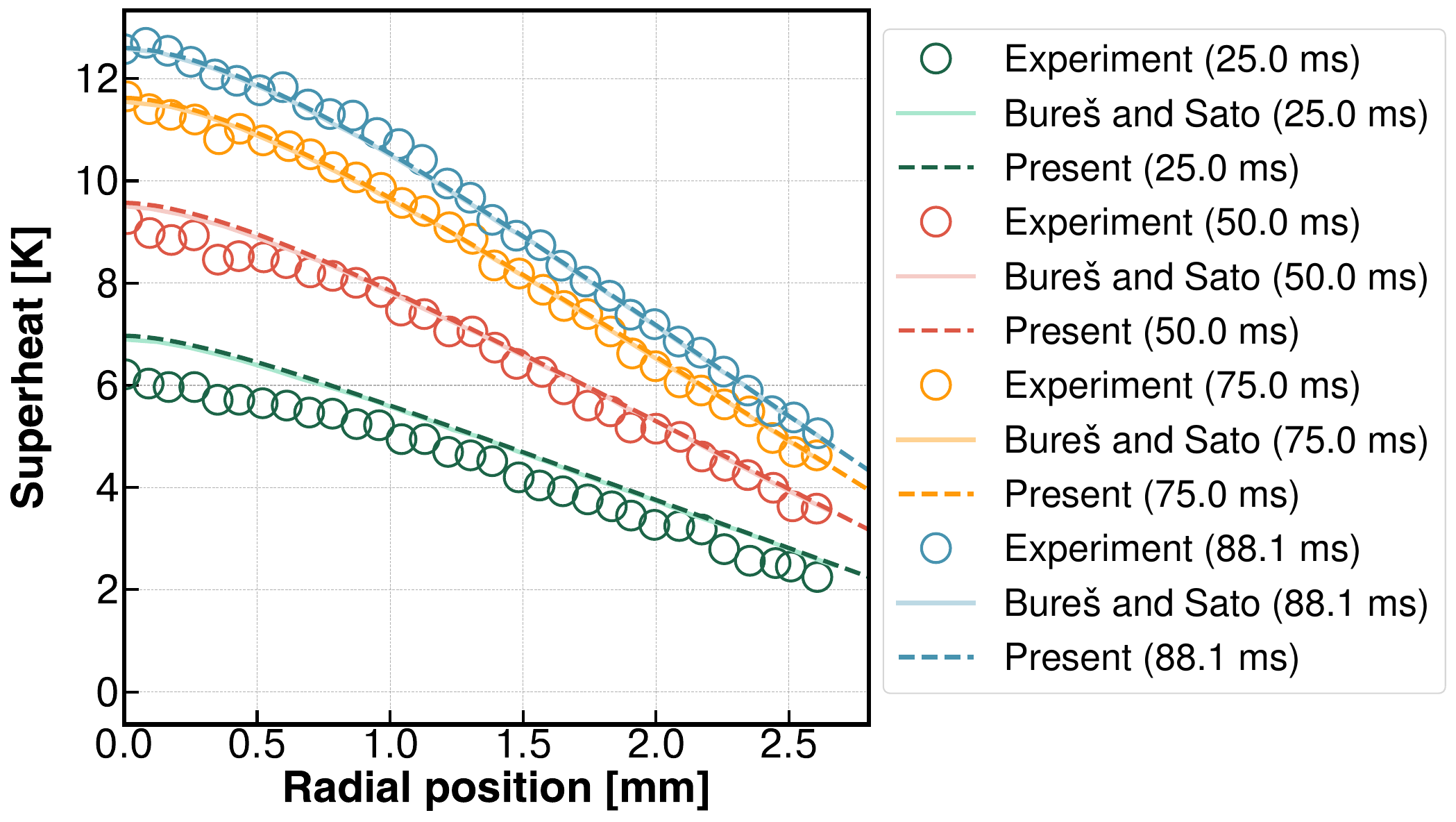}}
    \centering
    \caption{Evolution of the heater surface superheat before the onset of nucleation. The results of the present study are compared with the experimental data of \citet{bucci2020thesis} and the numerical results of \citet{burevs2022comprehensive}.}
    \label{Fig_initial_T}
\end{figure}
In the experiment, the system starts from saturated and stagnant conditions. After a period of heating, the first bubble nucleates in a cavity on the heated surface and then grows rapidly. Within the sharp-interface framework, an initial bubble seed is required to trigger the phase-change process. To model the nucleation process, following \citet{burevs2022comprehensive}, a transient heating problem without the vapor phase is solved until the temperature at the nucleation site reaches the nucleation temperature of $112.55^\circ$C. Given the nucleation superheat of $\Delta T = 12.55\ \rm{K}$, this problem can be characterized by the Jakob number:
\begin{equation}
    \mathrm{Ja} = \frac{\rho_{l}C_{p,l} \Delta T}{\rho_{v} h_{lg}}, 
\end{equation}
yielding a value of 37.5. Note that although the bubble growth results reported by \citet{bucci2020thesis} are axisymmetric, the heater geometry is rectangular. As a result, a uniform heat flux distribution under the axisymmetric numerical configuration leads to deviations in the calculated surface temperature distribution compared to experimental measurements. To address this issue, \citet{burevs2022comprehensive} adopted a modified heat flux distribution for the initial temperature field calculation, given as:
\begin{equation}
    j_{ini} = \mathrm{max}\left(\frac{4 - r}{4}\times 481, 0\right) (\rm{kW/m^2}).
\end{equation}
This linear power input is used to solve the liquid-solid heat transfer problem. To compute the initial temperature field, the domain is extended to $\left[ 0\ \rm{mm}, 7.0 \ \rm{mm} \right] \times \left[ -1.0\ \rm{mm}, 6.0 \ \rm{mm} \right]$. The simulations are performed with three grid levels ranging from 10 to 12, ensuring grid convergence (not shown here). In figure \ref{Fig_initial_T}, the superheat distribution on the solid surface at different time instants, obtained at grid level 12, is compared with experimental data \citep{bucci2020thesis} and the numerical results of \citet{burevs2022comprehensive}, showing excellent agreement. The waiting time (the time from the beginning of heating to nucleation) in the experiment ($88.1\ \rm{ms}$) is also well reproduced numerically.

\subsection{Numerical results}
\label{sec4.3:numerical_results}
\begin{figure}
     \centerline{\includegraphics[width=0.9\textwidth]{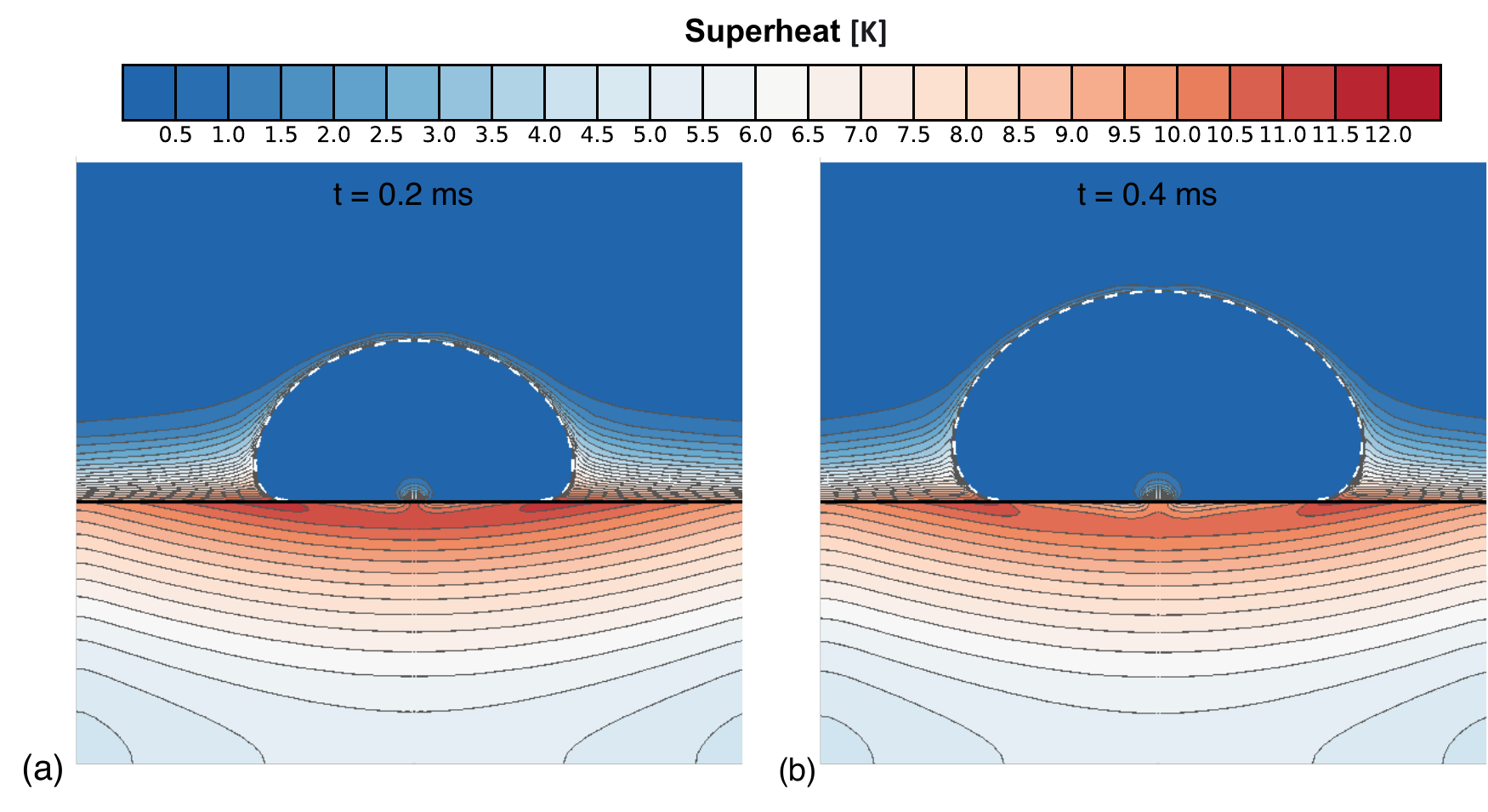}}
    \caption{The superheat distributions at (a) $t = 0.2\ \rm{ms}$  and (b) $t = 0.4\ \rm{ms}$ for case A ($\omega = 0.0345$), obtained at grid level 12. The white dashed line and the black solid line represent the liquid-vapor interface and the fluid-solid boundary, respectively. The results are mirrored about $r = 0$ for better visualization. }
    \label{Fig_whole_field}
\end{figure}
\begin{figure}
    \centerline{\includegraphics[width=0.7\textwidth]{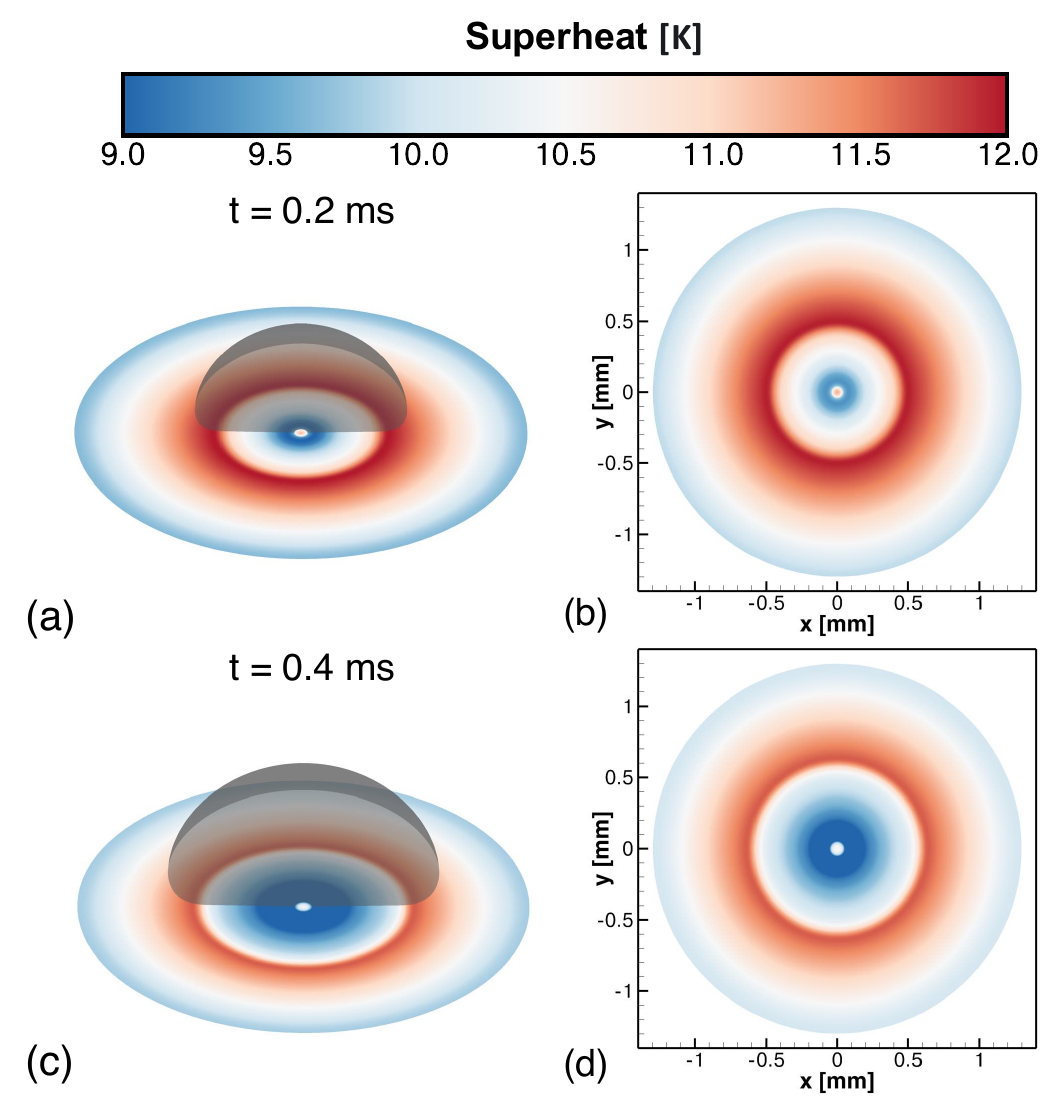}}
    \caption{The 3D bubble interfaces and the superheat distributions on the solid surface at different time instants for case A ($\omega = 0.0345$), obtained at grid level 12. The plots are generated by rotating the axisymmetric results.}
    \label{Fig_bubble_3D}
\end{figure}

Using the temperature distribution at $t = 88.1\ \rm{ms}$ as the initial condition, we position a bubble seed with a radius of $R_0 = 20\ \rm{\mu m}$ at the origin to initiate bubble growth. The simulations are performed up to $t = 0.5\ \rm{ms}$, during which a uniform heat flux of $425\ \rm{kW/m^2}$ is applied to the heater \citep{burevs2022comprehensive}. In this section, the numerical results obtained by the present model are validated against experimental data and previous numerical studies. Before the quantitative comparison, the superheat distributions within the computational domain at two time instants obtained at grid level 12 are shown in figure \ref{Fig_whole_field} for case A ($\omega = 0.0345$). As in previous studies \citep{burevs2022comprehensive, torres2024coupling}, the thermal boundary layer near the bubble surface becomes progressively thinner as the bubble grows. Additionally, the solid surface beneath the bubble is cooled due to vigorous evaporation within the microlayer. This phenomenon is more clearly observed in the 2D distribution of the superheat on the solid surface in figure \ref{Fig_bubble_3D}, which is generated by rotating the axisymmetric data. Notably, the highest superheat on the solid surface is located at the outer edge of the microlayer rather than at the lateral edge of the bubble.

\begin{figure}
    \centerline{\includegraphics[width=0.85\textwidth]{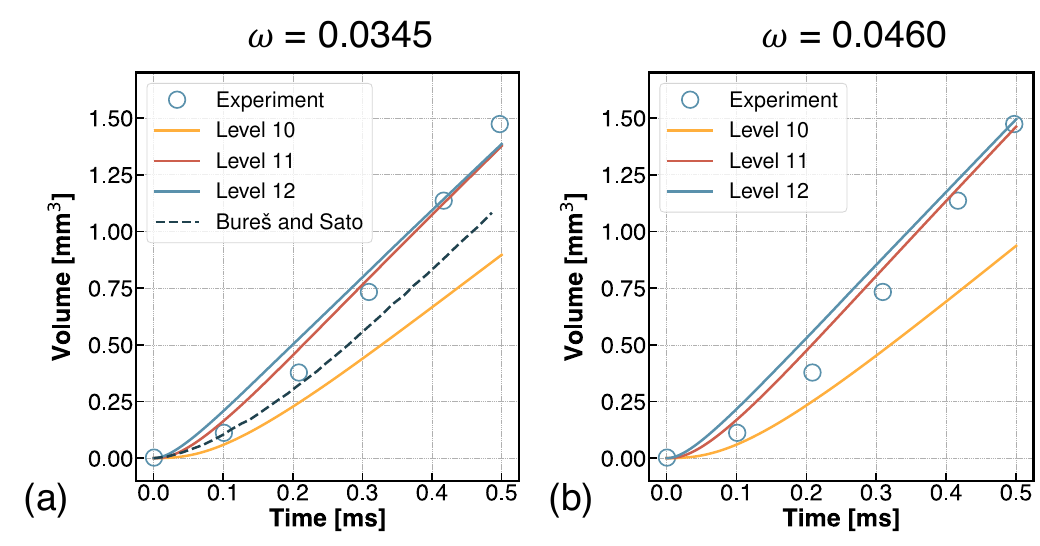}}
    \caption{The time histories of the bubble volume for (a) case A ($\omega = 0.0345$) and (b) case B ($\omega = 0.0460$) at different grid levels, compared with the experimental data of \citet{bucci2020thesis} and the numerical results of \citet{burevs2022comprehensive}. Note that the results for case B are not reported in the work of \citet{burevs2022comprehensive}.}
    \label{Fig_volume}
\end{figure}
\begin{figure}
    \centerline{\includegraphics[width=0.85\textwidth]{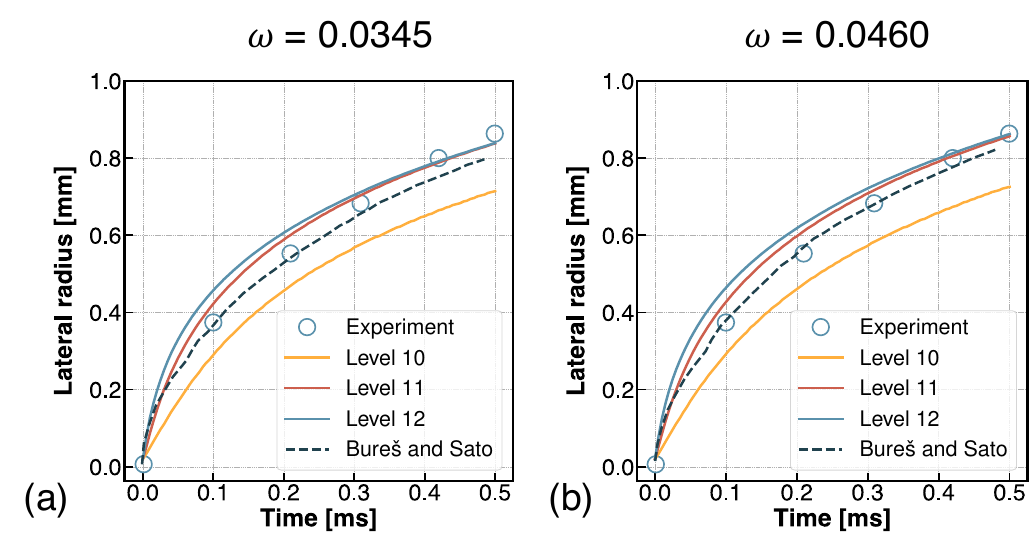}}
    \caption{The time histories of the bubble lateral radius for (a) case A ($\omega = 0.0345$) and (b) case B ($\omega = 0.0460$) at different grid levels, compared with the experimental data of \citet{bucci2020thesis} and the numerical results of \citet{burevs2022comprehensive}.}
    \label{Fig_radius}
\end{figure}
\begin{figure}
    \centerline{\includegraphics[width=0.9\textwidth]{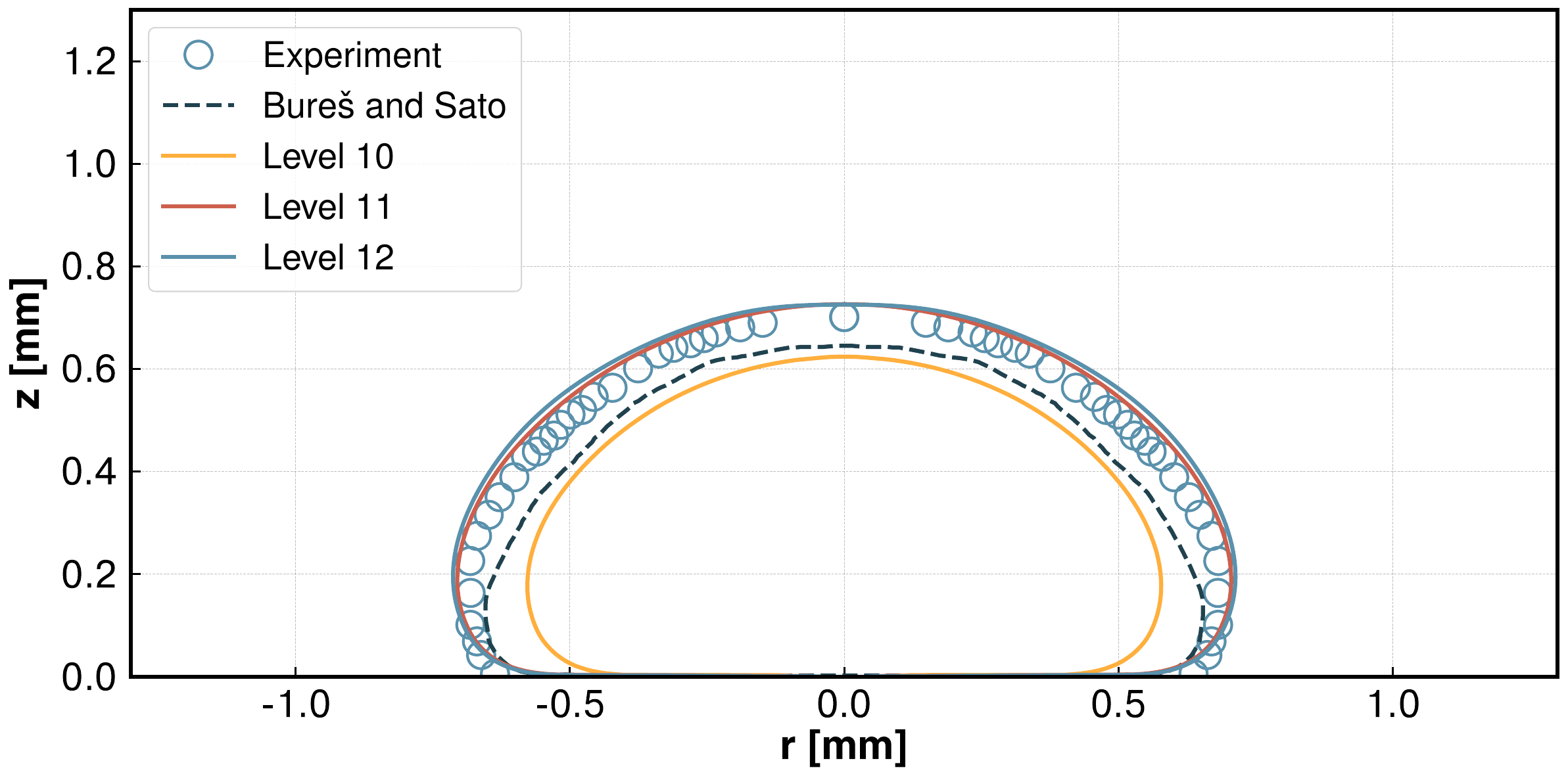}}
    \caption{The bubble interfaces at $t = 0.31\ \rm{ms}$ for case A ($\omega = 0.0345$) at different grid levels, compared with the experimental data of \citet{bucci2020thesis} and the numerical results of \citet{burevs2022comprehensive}. The results are mirrored about $r = 0$ for better visualization.}
    \label{Fig_shape_31}
\end{figure}
\begin{figure}
    \centerline{\includegraphics[width=1.0\textwidth]{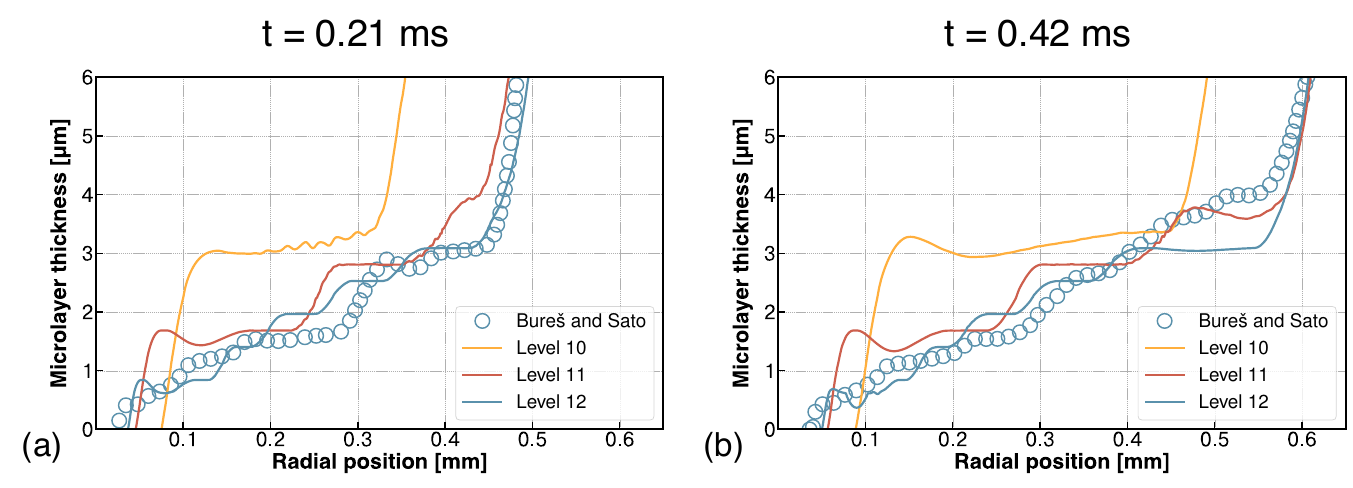}}
    \caption{The microlayer profiles at different time instants and grid levels for case A ($\omega = 0.0345$), compared with the numerical results of \citet{burevs2022comprehensive}.}
    \label{Fig_microlayer_thickness}
\end{figure}

We then turn to a quantitative comparison of growth characteristics. In figures \ref{Fig_volume} and \ref{Fig_radius}, the volume and lateral radius of the bubble are plotted against time for all grid resolutions, demonstrating grid convergence. It should be noted that grid convergence was not achieved in the results reported by \citet{burevs2022comprehensive}. This peculiar convergence behavior is probably caused by the one-fluid method used in their simulations, which, as shown in appendix, can lead to numerical oscillations within the microlayer due to strong heat transfer. In general, the convergent results obtained with the present method align well with the experimental data of \citet{bucci2020thesis} and the numerical results of \citet{burevs2022comprehensive} at their finest grid resolution with the smallest cell size of $0.475\ \rm{\mu m}$. For case A, the bubble volume and lateral radius are slightly overestimated before 0.3 ms and underestimated thereafter, compared to the experimental results. In contrast, for case B, the results show improved agreement in later stages. This is expected due to the higher accommodation coefficient in case B, which leads to lower IHTR and thus larger heat flux within the microlayer for the same superheat on the solid surface, according to equation (\ref{Eq_accomodation_coef}). The same trend is also observed in the results of \citet{burevs2022comprehensive}. In their simulations, for case A, the bubble volume and lateral radius are underestimated after 0.2 ms, whereas for case B, better agreement is observed in later stages. It is evident from figures \ref{Fig_volume} and \ref{Fig_radius} that the accommodation coefficient significantly influences the bubble growth. The discrepancies between the convergent numerical results and the experimental data may be attributed to the constant accommodation coefficient used throughout the simulations. In reality, a time-dependent accommodation coefficient, which can vary significantly from 0.01 to 1, has been observed in experiments \citep{marek2001analysis}. To our knowledge, while several theories exist \citep{nathanson1996dynamics, persad2016expressions}, accurately modeling the accommodation coefficient remains an open question in the literature.

A more detailed comparison of the bubble shape at $t = 0.31\ \rm{ms}$ is presented in figure \ref{Fig_shape_31}. For case A, the bubble interfaces at different grid levels are compared with the experimental data of \citet{bucci2020thesis} and the numerical result of \citet{burevs2022comprehensive}. In our results, the interfaces at grid levels 11 and 12 nearly overlap and agree well with the experimental data, demonstrating good grid convergence. Moreover, our results show a smoother bubble shape compared to the result of \citet{burevs2022comprehensive}, which is attributed to the reduced numerical oscillations achieved by the ghost fluid method. The bubble shape obtained by \citet{burevs2022comprehensive} is more flattened compared to both our results and the experimental data. This explains why their results are in good agreement with the experiment regarding the lateral radius, while the bubble volume is significantly underestimated. In addition to macroscopic bubble shapes, the microlayer profiles for case A at different time instants and grid levels are illustrated in figure \ref{Fig_microlayer_thickness}. The microlayer profile can be measured using laser interferometry \citep{jung2018hydrodynamic, narayan2021non}, though this technique was not utilized in Bucci's experiments \citep{bucci2020thesis}. Thus, we only compare our results with the numerical results of \citet{burevs2022comprehensive}, showing good agreement. The step-like profiles observed are numerical artifacts associated with the finite grid resolution, as also noted in previous studies \citep{burevs2022comprehensive, burevs2021modelling}. The result at grid level 10 for $t = 0.21\ \rm{ms}$ exhibits oscillations due to the coarse grid size ($2.246\ \rm{\mu m}$), which is insufficient to resolve the heat transfer within the microlayer. It is emphasized that the current setup is numerically challenging due to the involvement of multiple spatial scales and intense mass and heat transfer. In the work of \citet{burevs2022comprehensive}, strong spurious waves along the interface were reported. To prevent simulation crashes, they imposed very strict timestep limits ($C_{adv} = 0.02$ and $C_{\sigma} = 0.063$) and applied an artificial averaging procedure for the phase-change rate during the initial stage of bubble growth. In the present method, numerical stability is improved by the ghost fluid method, allowing the use of a larger timestep and eliminating the artificial averaging procedure.

\begin{figure}
    \centerline{\includegraphics[width=0.8\textwidth]{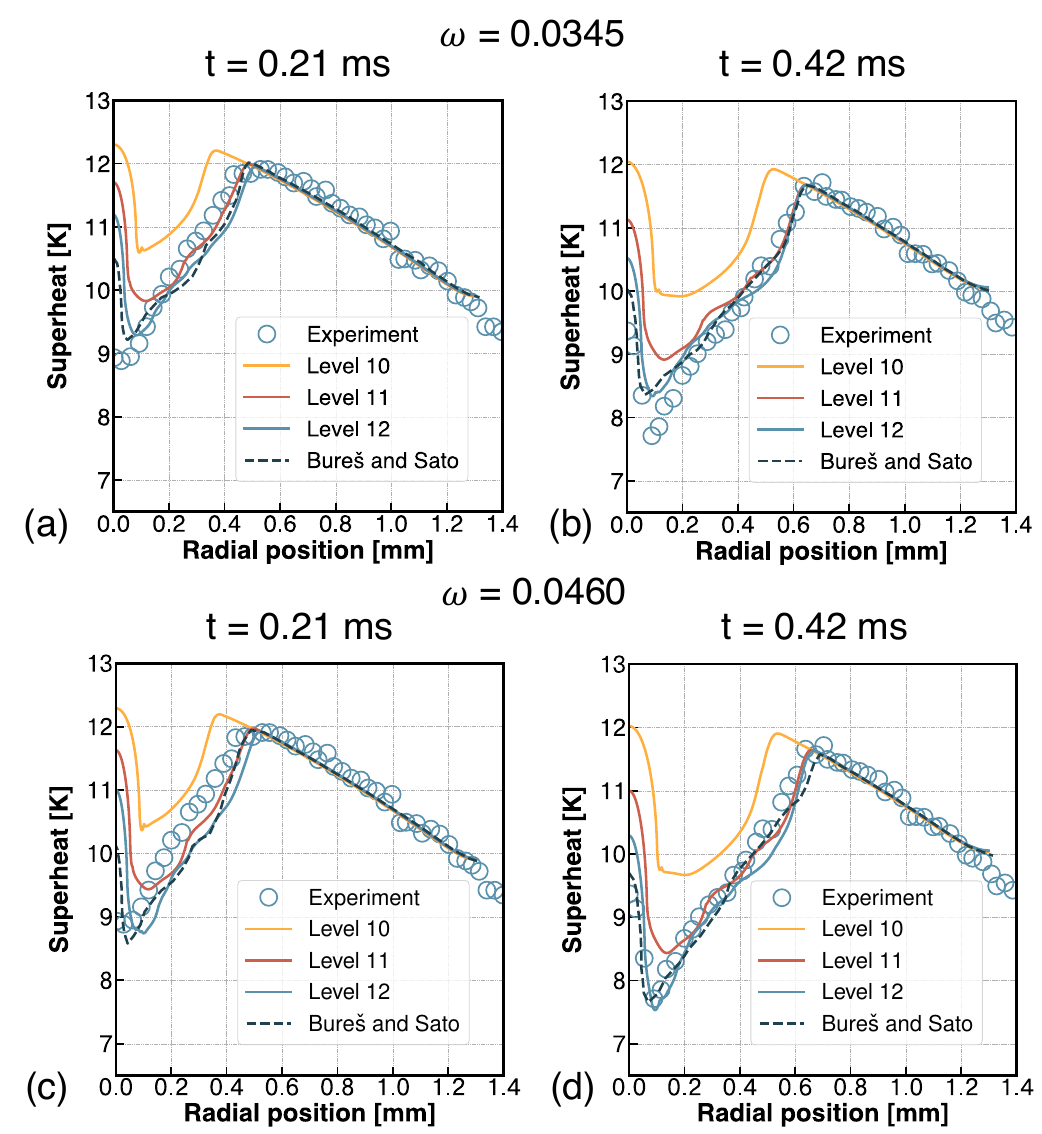}}
    \caption{The surface superheat distribution at different time instants and grid levels, compared with the experimental data of \citet{bucci2020thesis} and the numerical results of \citet{burevs2022comprehensive}. (a) and (b) are the results of case A ($\omega = 0.0345$), while (c) and (d) are the results of case B ($\omega = 0.0460$).}
    \label{Fig_superheat}
\end{figure}
\begin{figure}
    \centerline{\includegraphics[width=0.8\textwidth]{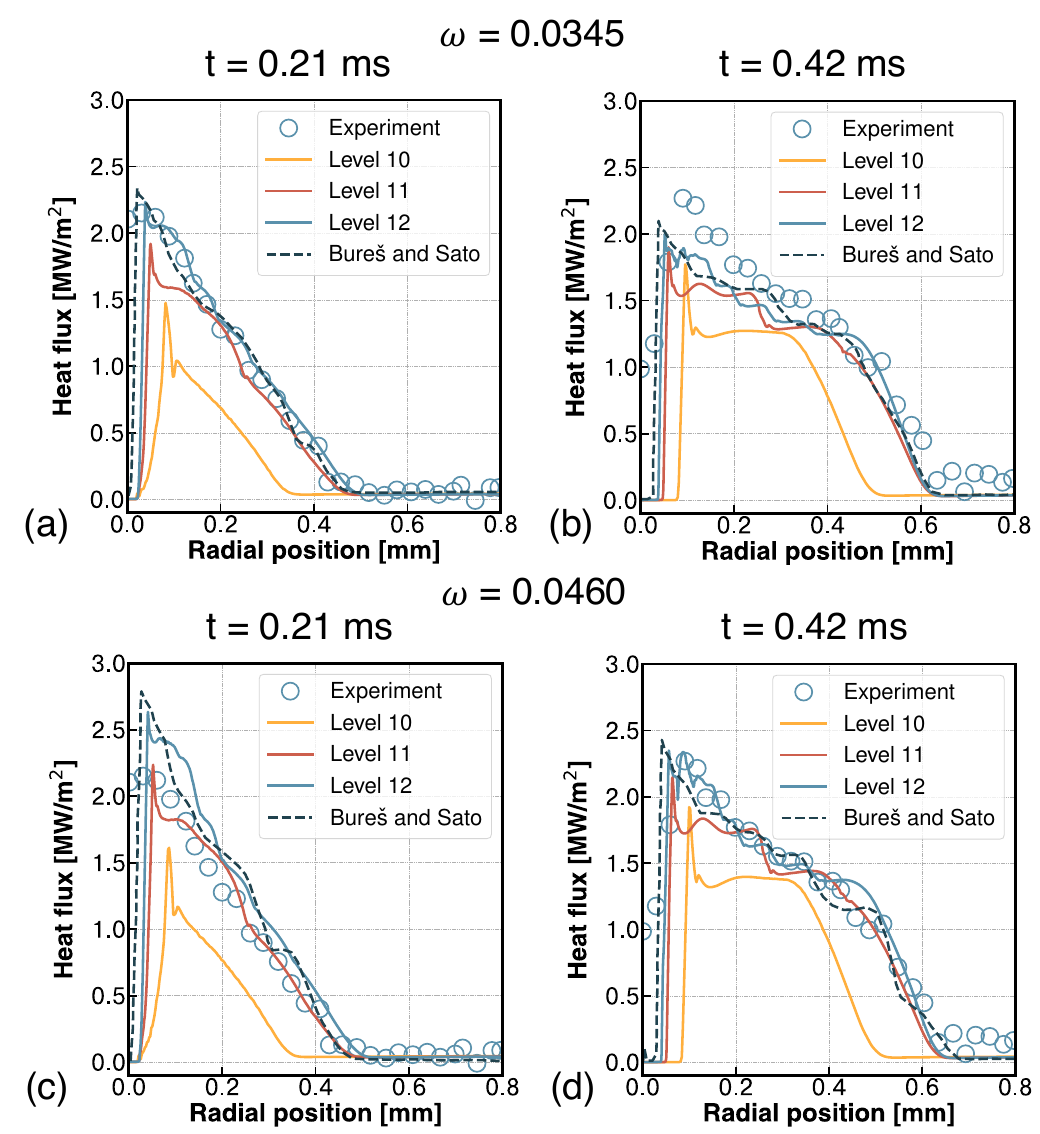}}
    \caption{The surface heat flux distribution at different time instants and grid levels, compared with the experimental data of \citet{bucci2020thesis} and the numerical results of \citet{burevs2022comprehensive}. (a) and (b) are the results of case A ($\omega = 0.0345$), while (c) and (d) are the results of case B ($\omega = 0.0460$).}
    \label{Fig_heat_flux}
\end{figure}

Finally, the thermodynamic characteristics for cases A and B are quantitatively validated. The surface superheat distributions at various time instants and grid levels are presented in figure \ref{Fig_superheat}. It is observed that as the grid resolution increases, our results progressively converge to the experimental data of \citet{bucci2020thesis}, showing excellent agreement with the numerical results of \citet{burevs2022comprehensive}. The extent of the microlayer can be identified from surface temperature variations: the temperature decreases from the origin to the contact line, increases along the microlayer, and then declines beyond the microlayer front. These results demonstrate that our method effectively captures the evolution of the microlayer. As in the work of \citet{burevs2022comprehensive}, larger deviations between numerical and experimental results are observed within the dry patch region (from the origin to the contact line). Note that detailed measurement uncertainties are not reported by \citet{bucci2020thesis}, and this discrepancy may be attributed to less accurate measurements on surfaces not covered by liquid \citep{burevs2022comprehensive}. Furthermore, the heat flux distributions at different times and resolutions are also measured and shown in figure \ref{Fig_heat_flux}. It can be seen that the heat flux peaks at the contact line and then decreases to zero along the radial extent of the microlayer. Notably, the heat flux within the microlayer region significantly exceeds the electrical power input. Before the bubble nucleation, the energy released from electrical resistance is distributed between both the liquid and solid. Given that the thermal diffusivity of the solid is considerably higher than that of the liquid, a greater proportion of energy is transferred to the solid. The subsequent cooling of the liquid due to boiling induces energy release from the solid, resulting in a substantially higher heat flux on the solid surface. This highlights the importance of including conjugate heat transfer between the fluid and solid in pool boiling simulations. It can be concluded from figure \ref{Fig_heat_flux} that the experimental measurements are well bounded by the simulation results of cases A and B, as observed by \citet{burevs2022comprehensive}. With a smaller accommodation coefficient (case A), better agreement is achieved in the early stages ($t = 0.21\ \rm{ms}$), while a larger accommodation coefficient (case B) yields improved results in the later stages ($t = 0.42\ \rm{ms}$). In the work of \citet{cai2024new}, a time-dependent accommodation coefficient model was developed and has been shown to improve the accuracy of IHTR models. In future work, we plan to improve the current model by incorporating such time-dependent models for the accommodation coefficient.

\subsection{Effect of contact angle}

\begin{figure}
    \centerline{\includegraphics[width=0.85\textwidth]{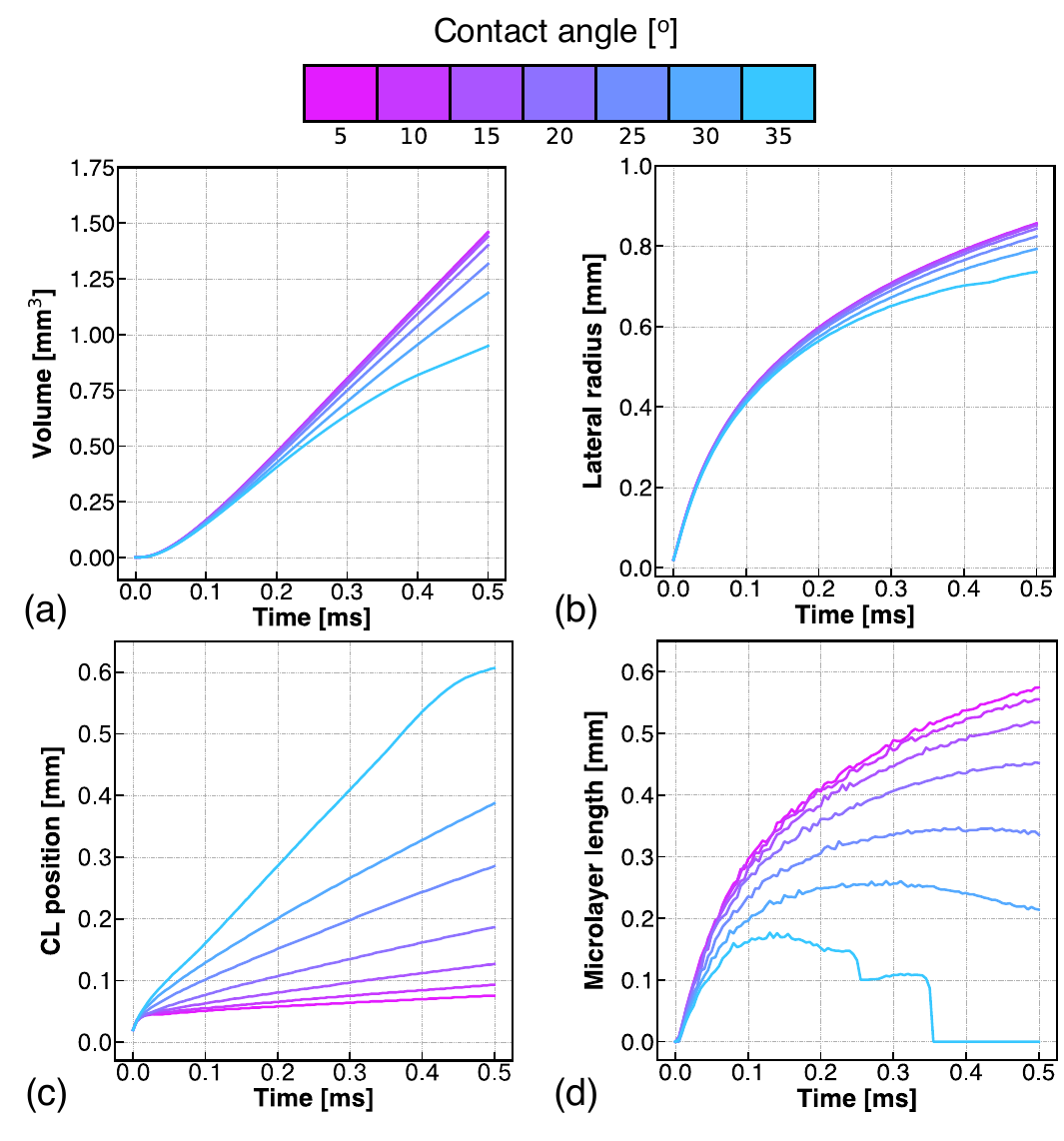}}
    \caption{The time histories of (a) the bubble volume, (b) the bubble lateral radius, (c) the position of the contact line, and (d) the microlayer length for various contact angles. The results are obtained with $\omega = 0.0460$ at grid level 11.}
    \label{Fig_data_CA}
\end{figure}
\begin{figure}
    \centerline{\includegraphics[width=1.0\textwidth]{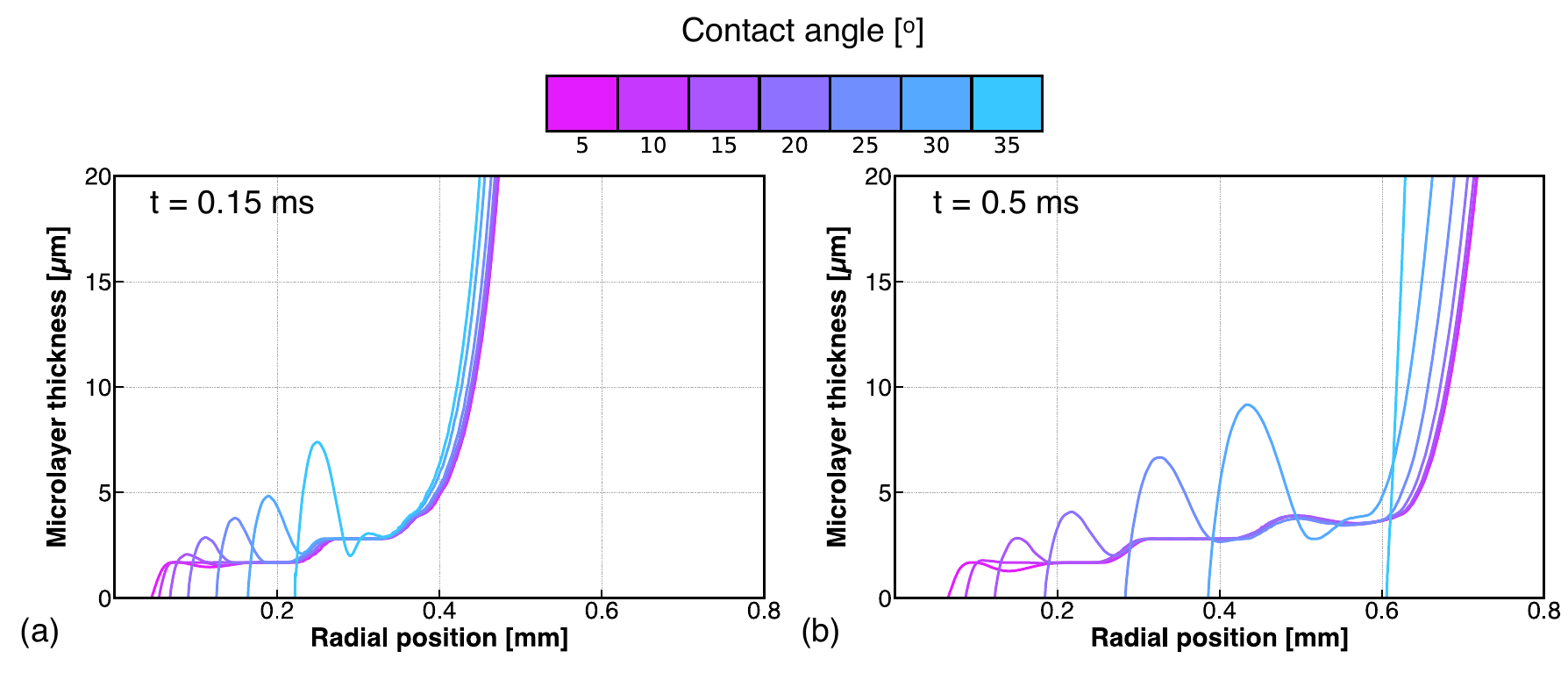}}
    \caption{The microlayer profiles at different time instants for various contact angles. The results are obtained with $\omega = 0.0460$ at grid level 11.}
    \label{Fig_ml_CA}
\end{figure}
\begin{figure}
    \centerline{\includegraphics[width=0.5\textwidth]{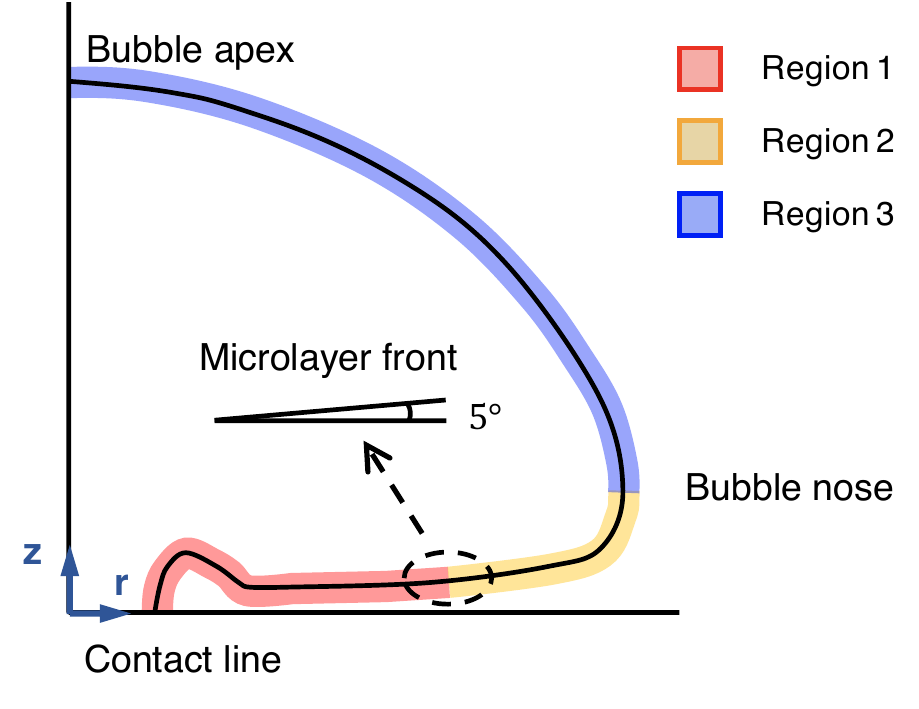}}
    \caption{Schematic of region partitioning. The region 1 extends from the contact line to the microlayer front. The region 2 spans from the microlayer front to the bubble nose, where the maximum width occurs. The region 3 extends from the bubble nose to the bubble apex.}
    \label{Fig_region_sketch}
\end{figure}

In the work of \citet{burevs2022comprehensive}, a dynamic contact angle model was proposed, where the predicted values of the contact angle throughout the simulation did not exceed $1^\circ$. As shown in section \ref{sec4.3:numerical_results}, our numerical results obtained with a static contact angle of $5^\circ$ are in very good agreement with those obtained using the dynamic contact angle model \citep{burevs2022comprehensive}. To further explore the influence of the contact angle, the same problem is simulated with varying contact angles $\theta_{C}$. Since it has been validated that grid level 11 is sufficient to capture the main physics of this problem, all simulations are performed at this grid level, with an accommodation coefficient of $\omega = 0.0460$. Seven contact angles, equally distributed within $[5^\circ, 35^\circ]$, are considered. 

We begin with a quantitative analysis of growth characteristics, focusing on the effect of the contact angle on microlayer development. In addition to the volume and lateral radius of the bubble, the contact line position $x_{CL}$ and the microlayer length $l_{ML}$ are also measured. The microlayer length is determined as the distance from the contact line to the microlayer front $x_{MF}$ \citep{urbano2018direct}, which is defined as the position where the interface slope exceeds $5^\circ$ (see figure \ref{Fig_region_sketch}). The results for various contact angles are presented in figure \ref{Fig_data_CA}. It is demonstrated that the influence of the contact angle on bubble growth diminishes as the angle decreases, particularly when it is less than $15^\circ$. As illustrated in figure \ref{Fig_data_CA}(c), after a short transition stage, the contact line moves at a constant velocity that is positively correlated with the contact angle. In particular, the results for $\theta_{C} = 35^\circ$ differ from others in the later stages due to the complete depletion of the microlayer, as indicated in figure \ref{Fig_data_CA}(d). Furthermore, a decrease in the growth velocity of the bubble lateral radius is observed over time, which corresponds to the bubble growth within the diffusion-controlled regime \citep{guion2018simulations}. As the bubble growth velocity decreases, the contact line velocity remains constant, and the extent of the microlayer is influenced by the competition between these two velocities. Simultaneously, evaporation contributes to microlayer depletion. At smaller contact angles, evaporation effects become more pronounced due to reduced contact-line mobility \citep{burevs2022comprehensive}, whereas dryout driven by hydrodynamic effects is more significant at larger contact angles \citep{urbano2018direct}. The microlayer profiles at various time instants for different contact angles are presented in figure \ref{Fig_ml_CA}. The transition from the microlayer regime to the contact line regime is observed for $\theta_{C} = 35^\circ$. Notably, at large contact angles, a dewetting ridge forms near the contact line, with its height increasing as the contact angle increases. This phenomenon results from mass conservation in the presence of enhanced contact-line mobility associated with larger contact angles \citep{giustini2024hydrodynamic}. Additionally, the central portions of the microlayer overlap for different contact angles, consistent with findings from previous studies \citep{guion2018simulations, giustini2024hydrodynamic}. In conclusion, different contact angles primarily influence interface motion near the contact line.

\begin{figure}[tbp]
    \centerline{\includegraphics[width=0.85\textwidth]{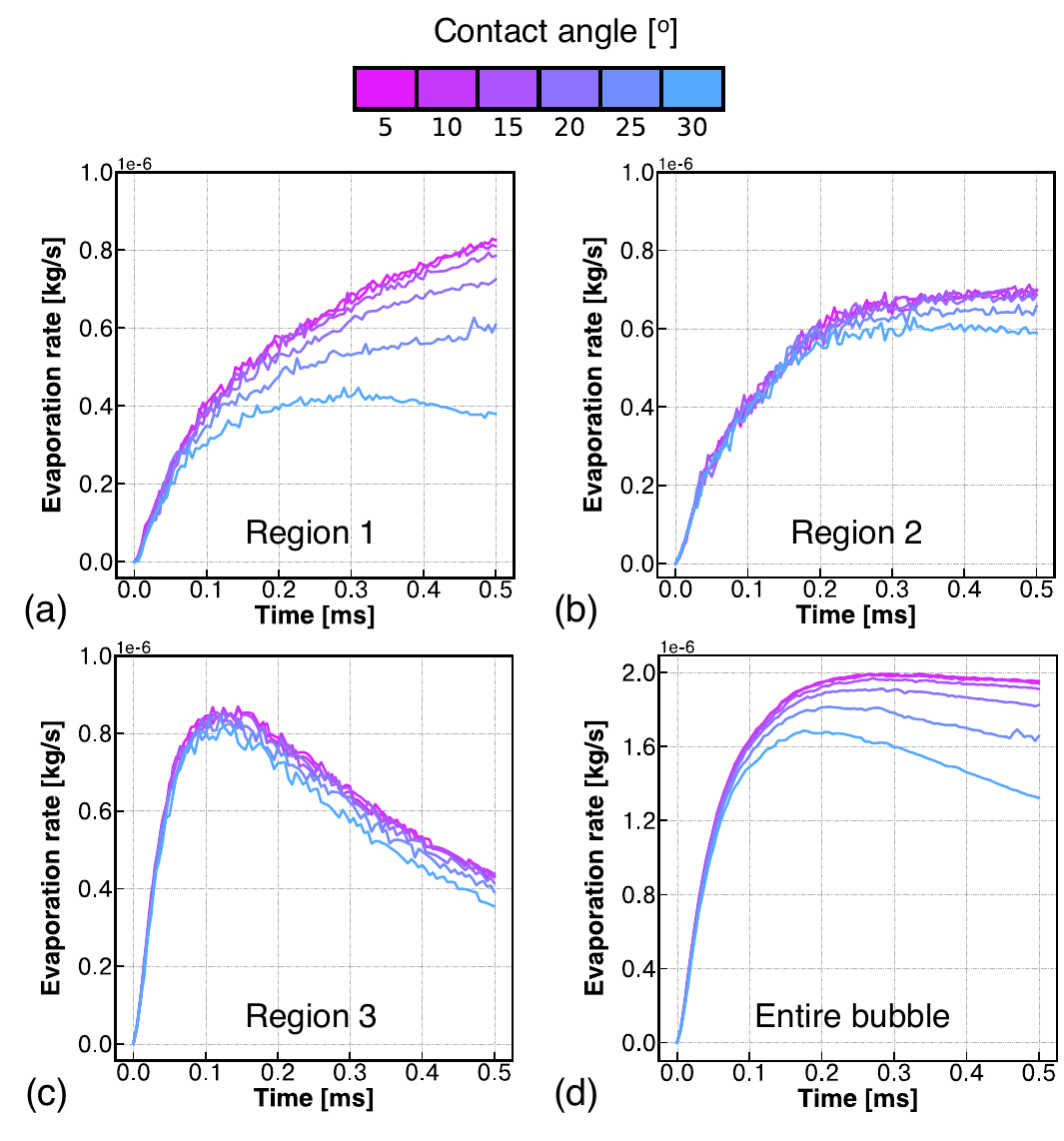}}
    \caption{The time histories of the evaporation rate over (a) the region 1, (b) the region 2, (c) the region 3, and (d) the entire bubble for various contact angles. The results are obtained with $\omega = 0.0460$ at level 11.}
    \label{Fig_mass_rate_CA}
\end{figure}

Subsequently, to thoroughly investigate the influence of the contact angle on thermodynamic characteristics, the bubble interface is divided into three regions, as illustrated in figure \ref{Fig_region_sketch}. Region 1, the microlayer region, extends from the contact line to the microlayer front. Region 2 spans from the microlayer front to the bubble nose, which is the farthest point along the radial direction. The remaining area, from the bubble nose to the bubble apex, is labeled as Region 3. It is important to note that the following analyses primarily focus on bubble growth in the microlayer regime. Consequently, the case of $\theta_C = 35^\circ$ is excluded, as it involves the transition from the microlayer regime to the contact line regime.

We begin the quantitative comparison by evaluating the evaporation rate, which is calculated as follows:
\begin{equation}
    r_e = \iint \dot{m} dA.
\end{equation}
The time histories of the evaporation rate within different regions are presented in figure \ref{Fig_mass_rate_CA}. It is observed that, generally, larger contact angles result in smaller evaporation rates, a trend validated across all three regions. In region 1, the results vary more significantly with different contact angles, while in regions 2 and 3, the results are more consistent across various contact angles. This phenomenon is expected since the evaporation rate is related to the integral area, and the microlayer extent is more sensitive to variations in the contact angle compared to the bubble volume and lateral radius. Following the trend of the microlayer length shown in figure \ref{Fig_data_CA}(d), the evaporation rate in region 1 increases at a decreasing rate over time. Notably, with a contact angle of $30^\circ$, the evaporation rate in region 1 begins to decrease after $t = 0.3$ ms, as the velocity of the microlayer front falls below that of the contact line. In region 2, the evaporation rate first increases and then stabilizes at $t = 0.2$ ms. The situation in region 3 is noticeably different: the evaporation rate initially increases, peaks at $t = 0.13$ ms, and then decreases rapidly. The different trends in the three regions arise because regions 1 and 2 are close to the solid surface and remain within the thermal boundary layer. In contrast, region 3, the upper surface region of the bubble, moves above the thermal boundary layer due to its axial growth, as observed in figure \ref{Fig_whole_field}.

\begin{figure}
    \centerline{\includegraphics[width=0.85\textwidth]{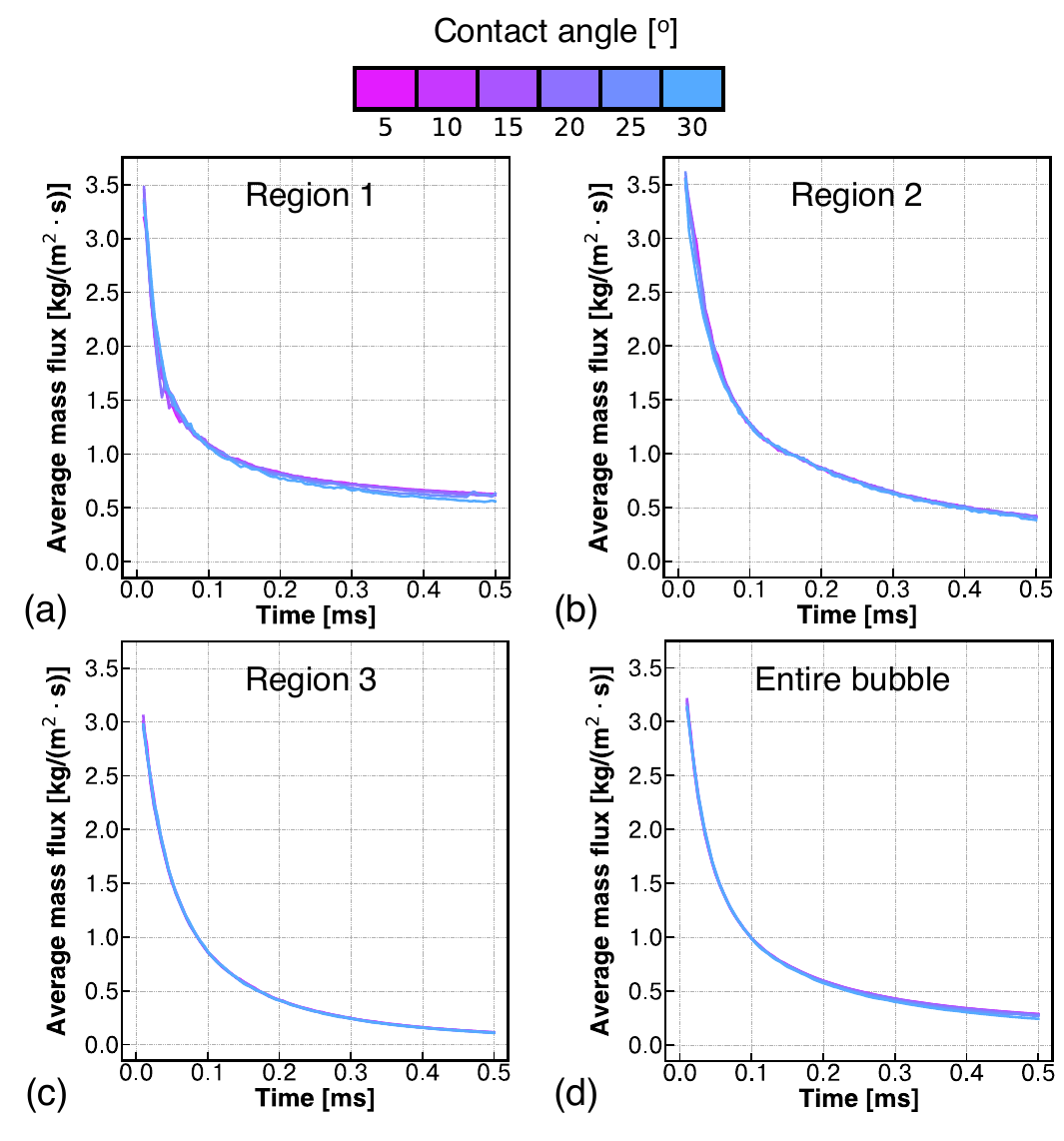}}
    \caption{The time histories of the average mass flux over (a) the region 1, (b) the region 2, (c) the region 3, and (d) the entire bubble for various contact angles. The results are obtained with $\omega = 0.0460$ at grid level 11.}
    \label{Fig_mass_flux_CA}
\end{figure}
\begin{figure}
    \centerline{\includegraphics[width=0.85\textwidth]{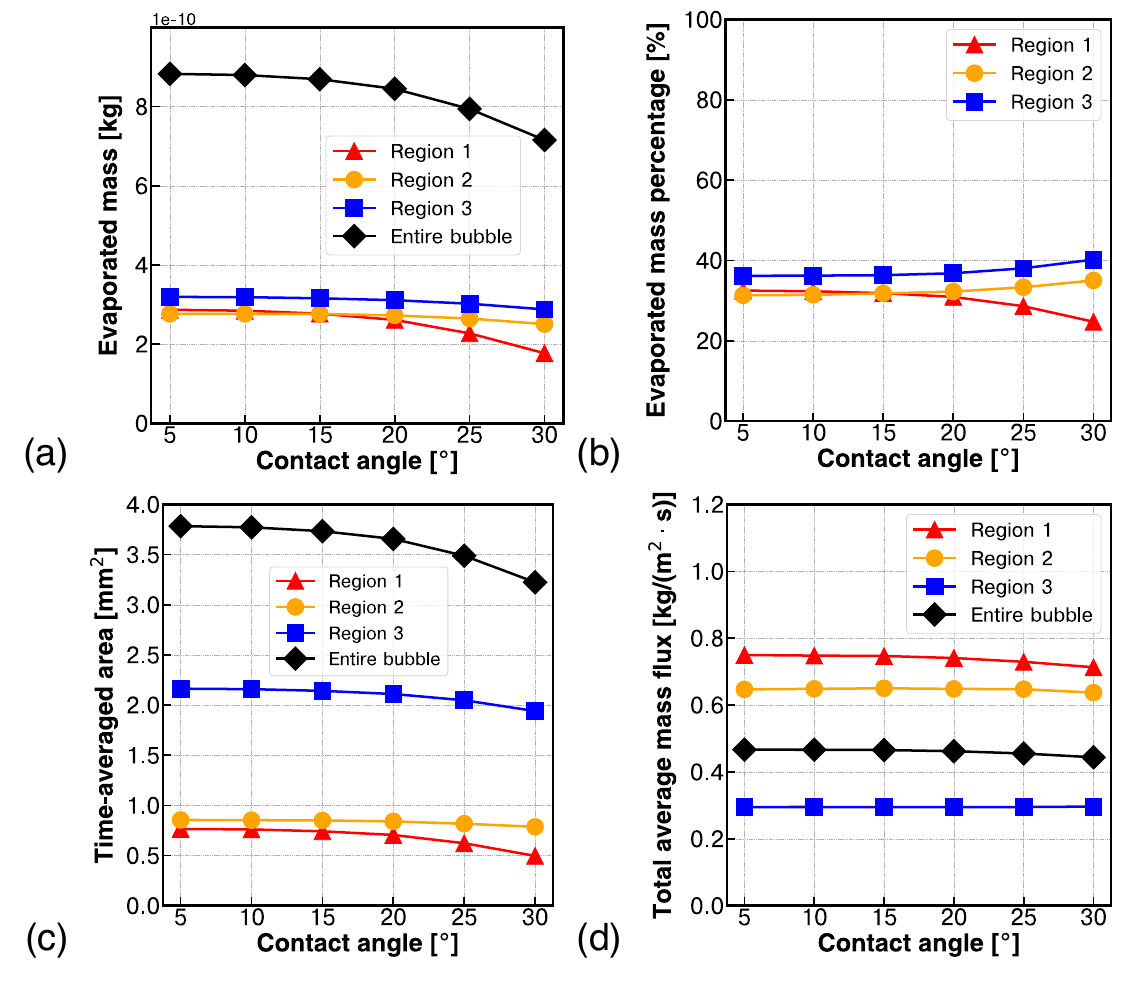}}
    \caption{(a) The total evaporated mass in different regions for different contact angles. (b) The percentage of the evaporated mass in each region relative to the total evaporated mass for different various angles. (c) The time-averaged surface area for different regions and contact angles. (d) The total average mass flux in different regions for different contact angles. The results are obtained with $\omega = 0.0460$ at grid level 11.}
    \label{Fig_average_CA}
\end{figure}

As the evaporation rate is strongly related to the integral area, the average mass flux, defined as
\begin{equation}
    \widetilde{\dot{m}} = \frac{\iint \dot{m} dA}{\iint dA},
\end{equation}
is considered to further investigate the influence of the contact angle. The time histories of the average mass flux within different regions are presented in figure \ref{Fig_mass_flux_CA}. It is observed that, for all regions, the average mass flux decreases at a decelerating rate, consistent with the theoretical analysis \citep{scriven1959dynamics} of bubble growth driven by heat diffusion. Notably, as shown in figure \ref{Fig_mass_flux_CA}, the average mass flux for different contact angles matches closely, especially in regions 2 and 3. This suggests that the observed differences in evaporation rates are primarily due to variations in integral area. 

Additionally, beyond the temporal variations of thermal characteristics, the cumulative behavior is investigated. The total evaporated mass $M_e$ is calculated by integrating the evaporation rate over time:
\begin{equation}
    M_e = \int_0^{t_f}\left(\iint \dot{m} dA\right)dt,
\end{equation}
where $t_f = 0.5\ \rm{ms}$ is the final time of the simulations. For different contact angles, the total evaporated mass in different regions is computed and shown in figure \ref{Fig_average_CA}(a). It can be seen that as the contact angle increases, the evaporated mass in the microlayer region is most affected. The proportions of the evaporated mass in each region relative to the total evaporated mass for different contact angles are detailed in figure \ref{Fig_average_CA}(b). As the contact angle increases from $5^\circ$ to $30^\circ$, the contribution from the microlayer region decreases from 31\% to 25\%. In figure \ref{Fig_average_CA}(c), the time-averaged surface area in different regions,
\begin{equation}
    \overline{S} = \frac{1}{t_f}\int_0^{t_f}\left(\iint dA\right)dt,
\end{equation}
is evaluated for different contact angles. The correlation between the trends of the average surface area and the evaporated mass relative to the contact angle is evident. To specifically investigate the influence of the contact angle on global thermal effects, we define the total average mass flux as:
\begin{equation}
    \overline{\dot{m}} = \frac{M_e}{\overline{S}\ t_f} = \frac{\int_0^{t_f}\left(\iint \dot{m} dA\right)dt}{\int_0^{t_f}\left(\iint dA\right)dt}. 
\end{equation}

The results for different regions are plotted against the contact angle in figure \ref{Fig_average_CA}(d). It is observed that, although the evaporated mass and surface area vary with different contact angles, the total average mass flux remains consistent. In particular, in regions 2 and 3, the total average mass fluxes are almost constant, regardless of the contact angle. As the contact angle increases from $5^\circ$ to $30^\circ$, the total average mass flux over the entire bubble decreases slightly (from $0.47\ \rm{kg/(m^2\cdot s)}$ to $0.44\ \rm{kg/(m^2\cdot s)}$), mainly due to minor changes in the microlayer thickness. Overall, it can be concluded that the hydrodynamic effect is the dominant factor influencing bubble growth over different contact angles, while the thermal effect remains consistent regardless of the contact angle. A larger contact angle negatively affects the evaporation process because the increased mobility of the contact line results in a smaller surface area within the microlayer region. Over time, the reduction of area in the microlayer region, region 1, slows down the bubble expansion due to the reduced amount of vapor evaporated from this region, yielding a smaller area for regions 2, and 3. The reduction of the evaporation rate in regions 2 and 3, in turn, further decelerates bubble growth. This explains why the differences in results for different contact angles increase with time, as shown in figure \ref{Fig_data_CA}. 

\subsection{Complete bubble cycle}
\label{sec4.6:complete_bubble_cycle}
\begin{table}
\begin{center}
\def~{\hphantom{0}}
\begin{tabular}{c c c c}
&Level 11 & Level 12 & Level 13 \\
\hline Minimum grid size $[\rm{\mu m}]$& 2.69 & 1.34 & 0.67 \\
Number of cores $[-]$& 24 & 48 & 64 \\
CPU time [core-h]&$\sim500$ & $\sim4900$ & $\sim17000$ \\
\end{tabular}
\caption{Computational cost for the DNS of a complete bubble cycle.}
\label{Tab_cost_complete}
\end{center}
\end{table}
\begin{figure}
    \centerline{\includegraphics[width=0.85\textwidth]{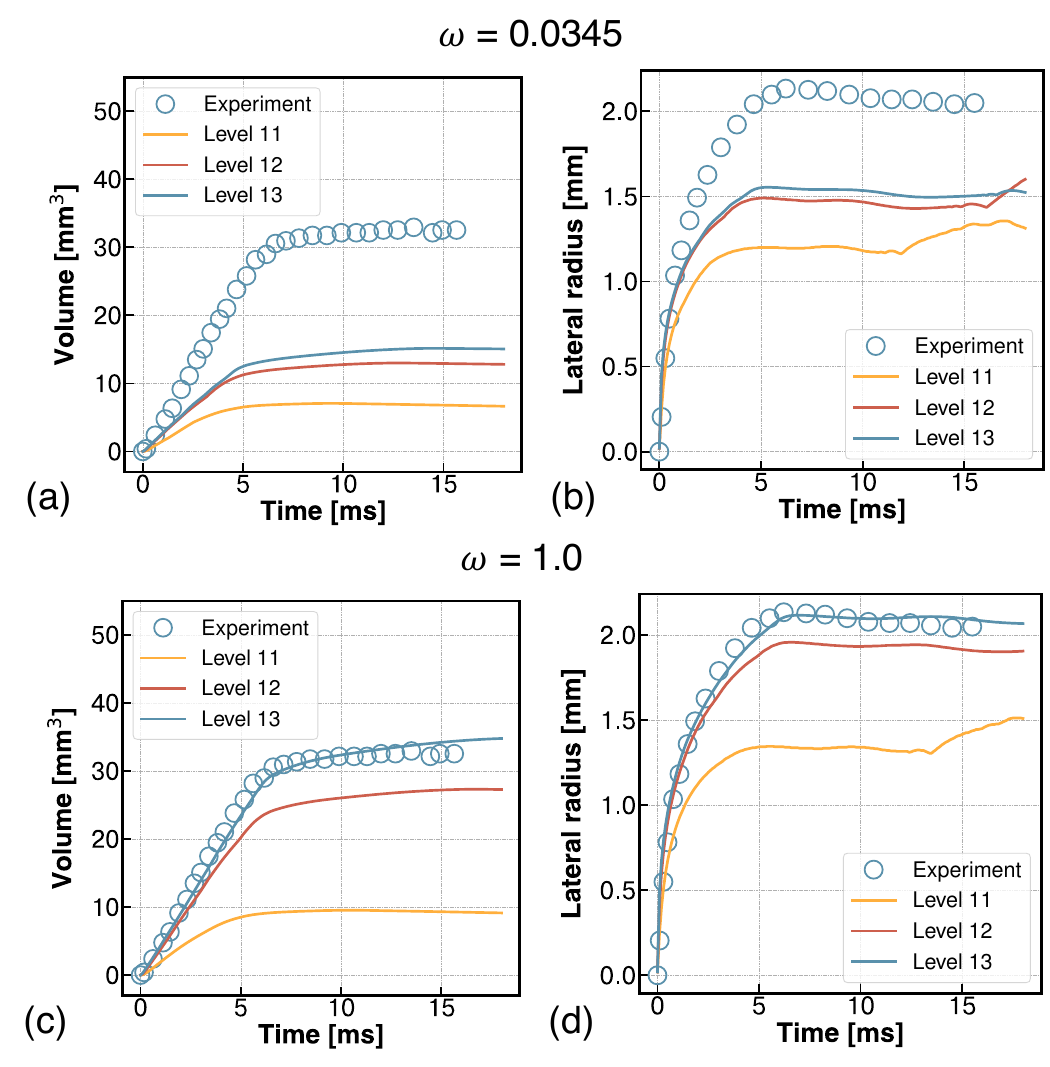}}
    \caption{The time histories of the bubble volume and the bubble lateral radius at different grid levels. The simulation results obtained with $\omega = 0.0345$ ((a) and (b)) and $\omega = 1.0$ ((c) and (d)) are compared with the experimental data of \citet{bucci2020thesis}.}
    \label{Fig_long_growth}
\end{figure}
\begin{figure}
    \centerline{\includegraphics[width=0.85\textwidth]{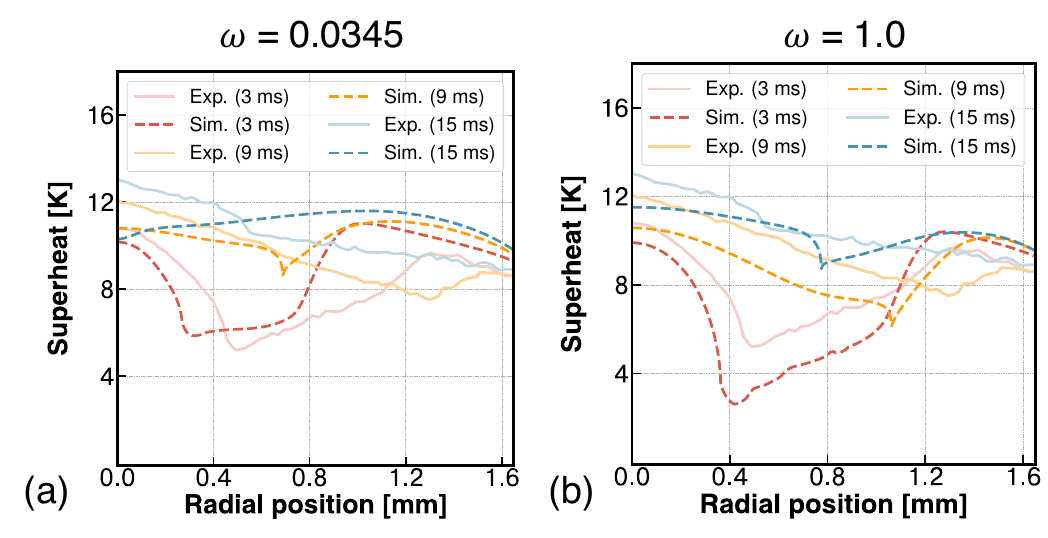}}
    \caption{The superheat distributions at grid level 13 for different time instants. The simulation results obtained with (a) $\omega = 0.0345$ and (b) $\omega = 1.0$ are compared with the experimental data of \citet{bucci2020thesis}.}
    \label{Fig_long_superheat}
\end{figure}
\begin{figure}
    \centerline{\includegraphics[width=1.0\textwidth]{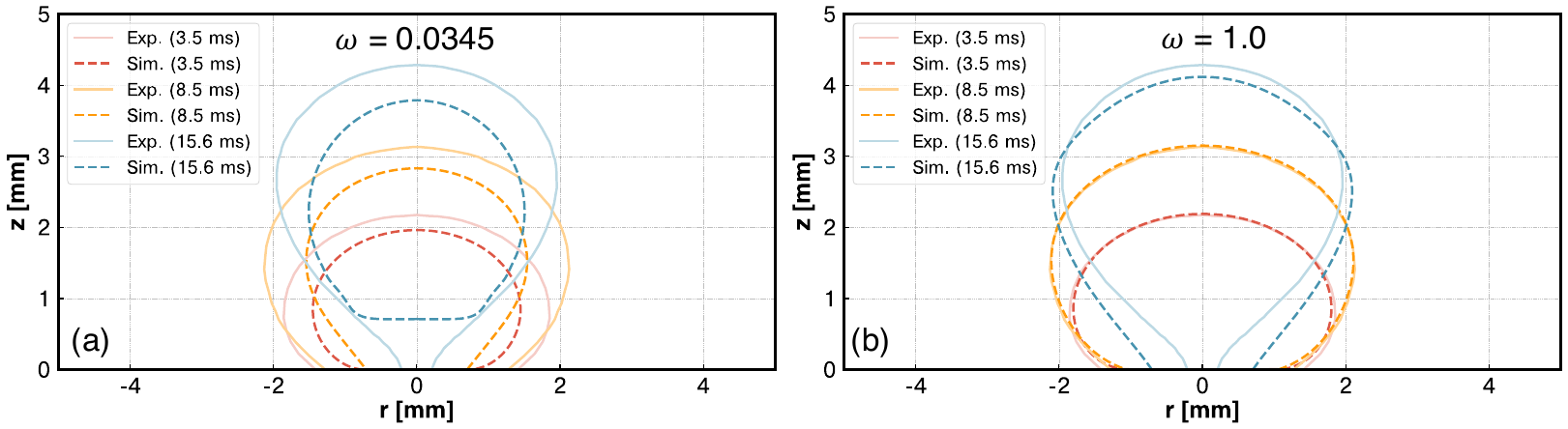}}
    \caption{The bubble interfaces at grid level 13 for different time instants. The simulation results obtained with (a) $\omega = 0.0345$ and (b) $\omega = 1.0$ are compared with the experimental data of \citet{bucci2020thesis}. The results are mirrored about $r = 0$ for better visualization.}
    \label{Fig_long_interfaces}
\end{figure}

In previous DNS studies of nucleate boiling in the microlayer regime \citep{burevs2022comprehensive, urbano2018direct}, the total physical time simulated was generally quite short (less than 2 ms) due to high computational costs. This duration is notably brief compared to the entire bubble cycle from nucleation to detachment. In this section, using the present, more efficient and stable solver, we perform DNS for the entire cycle of bubble growth, which, to our knowledge, is the first reported in the open literature. The setup remains consistent with the previous configuration, as shown in figure \ref{Fig_sketch_domain}, except that the computational domain is extended to $\left[ 0\ \rm{mm}, 5.5 \ \rm{mm} \right] \times \left[ -1.0\ \rm{mm}, 4.5 \ \rm{mm} \right]$ to capture the larger bubble encountered during the simulation. Three grid levels, from 11 to 13, are adopted, resulting in minimum grid sizes ranging from $2.69\ \rm{\mu m}$ to $0.67\ \rm{\mu m}$. The computational costs for simulations up to the physical time of 18 ms are detailed in table \ref{Tab_cost_complete} for different grid resolutions. The promising computational efficiency demonstrates the potential applications of the present solver in studies of more challenging nucleate boiling problems, such as flow boiling. 

In the experiment of \citet{bucci2020thesis}, after the depletion of the microlayer, contact angle hysteresis is observed. The contact angle hysteresis effect is that a contact line can recede (i.e., expansion of the dry region) only for contact angles lower than a critical value $\theta_{rec}$ and advance (i.e., shrinkage of the dry region) for contact angles higher than another critical value $\theta_{adv}$. These two critical angles are referred to as the receding contact angle and the advancing contact angle. Due to the presence of the microlayer, the receding contact angle cannot be measured, while the advancing contact angle is measured at approximately $50^\circ - 55^\circ$. In the present work, a contact-angle hysteresis model \citep{bures2024coarse, fang20083Dcontactline} is adopted, and the contact angle is updated from time step $n$ to $n+1$ by
\begin{equation}
    \theta^{n+1}=\max \left[\theta_{rec }, \min \left(\theta_{adv}, \theta^n-2 \sin ^2\left(\theta^n\right) \frac{H_l^{n+1}-H_l^n}{\Delta}\right)\right],
    \label{Eq_hysteresis}
\end{equation}
where $H_l$ represents the total wall-parallel liquid height in the contact-line cell row. This formula is derived based on the assumption that, for a contact angle between the receding angle $\theta_{rec}$ and the advancing angle $\theta_{adv}$, any change in the liquid content within the contact-line cell row corresponds to a rotation of the interface rather than a shift in the contact-line position. Following \citet{bures2024coarse}, we use $\theta_{rec} = 5^\circ$ and $\theta_{adv} = 55^\circ$ in our simulations.

Subsequently, the numerical results are compared with the experimental data. With $\omega = 0.0345$, the time histories of the bubble volume and the bubble lateral radius at different grid levels are presented in figures \ref{Fig_long_growth}(a) and (b). It can be seen that the bubble growth rate is underestimated after $t = 2.0$ ms, although good agreement is observed in the early stages. The oscillations of the lateral radius in the later stages are caused by deformations of the bubble after its detachment. In addition to $\omega = 0.0345$, the other accommodation coefficient used by \citet{burevs2022comprehensive}, $\omega = 0.0460$, still leads to a significant underestimation of the bubble growth rate, and the corresponding results are not shown here for brevity. It is important to note that the modeling of the accommodation coefficient remains an open question in the literature \citep{cai2024new}. Our goal here is not to solve this problem but to demonstrate the effect of the accommodation coefficient on the DNS of complete bubble cycles in nucleate boiling. In previous studies \citep{burevs2022comprehensive, el2024numerical, torres2024coupling}, the physical time of the simulations was not large enough to observe these differences. In the present work, the perfect evaporation case ($\omega = 1.0$) is also simulated, and the results are shown in figures \ref{Fig_long_growth}(c) and (d). It can be observed that, with perfect evaporation within the microlayer, the growth of the bubble volume and the bubble lateral radius is in better agreement with the experimental measurements. This is expected because diminishing interfacial heat-transfer resistance, corresponding to an increasing accommodation coefficient, leads to a higher evaporation rate. 

In figure \ref{Fig_long_superheat}, the superheat distributions along the wall at different time instants are compared with the experimental data. With $\omega = 0.0345$, at $t = 3$ ms and $t = 9$ ms, despite the deviations in the positions of the contact line and the microlayer front (caused by the underestimated bubble growth rate), the predicted lowest superheat on the wall is in better agreement with the experiment compared to that in the perfect evaporation case. A possible explanation for these deviations is that the rectangular heater is modeled as an axisymmetric configuration in numerical simulations. Consequently, the energy input is lower than in actual experiments due to the missing heater area. This accumulated effect might not have been identified in previous studies due to the limited simulation time. In the perfect evaporation case, this energy loss is compensated by a lower IHTR, allowing more energy transfer from the solid to the fluid. However, in the current case, the selected accommodation coefficient overcompensates for this loss, leading to an underestimation of the superheat on the solid wall. These findings highlight the necessity of 3D simulations to better assess the cumulative effect of heater geometry. The highly efficient solver presented in this paper provides a promising platform for tackling such challenging simulations in future work. At $t = 15$ ms, the superheat distribution with $\omega = 0.0345$ deviates more from the experimental data than that with $\omega = 1.0$. This deviation is due to the advanced detachment of the bubble, as shown in figure \ref{Fig_long_interfaces}, where the bubble interfaces at different time instants are presented. It can be seen that at $t = 3$ ms and $t = 9$ ms, the interfaces obtained with $\omega = 1.0$ are in excellent agreement with the experimental measurements. A larger deviation is observed at $t = 15.6$ ms, which could be attributed to the contact angle model. During the detachment stage, according to Equation (\ref{Eq_hysteresis}), the contact angle in the simulations is fixed at $\theta_C = \theta_{adv} = 55^\circ$. The time-dependent contact angle observed during the detachment stage in the experiment \citep{bucci2020thesis} is not modeled in the present study. Hence, it can be concluded that a better dynamic contact angle model is needed for accurately predicting the detachment at the end of the bubble cycle. This has not been discussed in previous DNS studies \citep{burevs2022comprehensive, urbano2018direct, torres2024coupling, el2024numerical}, as the simulation time was not long enough to reach the detachment stage.

\section{Conclusion}
\label{sec5:conclusion}

We have extended the open-source phase-change model developed by \citet{cipriano2024multicomponent} to simulate nucleate boiling in the microlayer regime with resolved conjugate heat transfer. Following \citet{burevs2022comprehensive}, heat transfer in the fluid and solid is coupled in a fully implicit manner, with a temperature jump condition accounting for the IHTR. The current work is based on the free software Basilisk \citep{popinet2009accurate,popinet2015quadtree}, in which the quad/octree-based AMR technique is employed to improve computational efficiency. To facilitate the implementation of AMR, a cell-centered velocity is adopted, though it can only be approximately projected (equation (\ref{Eq_discretization_5})). Based on the approximate projection method, the original model of \citet{cipriano2024multicomponent} works well for numerous benchmark tests \citep{cipriano2024multicomponent}. However, we have shown in appendix \ref{appA:ghost_fluid_method_compare} that the original model of \citet{cipriano2024multicomponent} introduces significant numerical oscillations within the microlayer \citep{zhao2022boiling, long2024edge}, thereby failing to accurately predict its development. The ghost fluid method, which removes the singularity at the interface by setting the ghost velocity, is employed and effectively suppresses these oscillations. With the ghost fluid method, we have successfully replicated the pool boiling experiments conducted at MIT \citep{bucci2020thesis}. The numerical results are in good agreement with the experimental data of \citet{bucci2020thesis} and the previous numerical results of \citet{burevs2022comprehensive}. With the help of AMR, computational efficiency is significantly improved, and the required CPU hours are reduced by three orders of magnitude. We thus believe the current work makes the present model more applicable for complex phase-change problems with high fidelity. The codes and configurations for all simulations are released in the Basilisk sandbox \citep{tiansandbox}.

Subsequently, with the present model, the influence of the contact angle on nucleate boiling in the microlayer regime is investigated. We have shown that the value of the contact angle influences the results in a decelerating manner, and very small contact angles, such as those predicted by the dynamic contact angle model proposed by \citet{burevs2022comprehensive}, are not necessary for the current scenario. Moreover, by dividing the bubble surface into three regions, we have shown that the influence of the contact angle is mainly confined to the microlayer region. It is demonstrated that thermal effects exhibit similarity across different contact angles, while hydrodynamic effects predominantly influence bubble growth. As the contact angle increases, the growing contact line mobility leads to a smaller surface area, while the total average mass flux remains approximately constant. Moreover, a complete bubble cycle from nucleation to detachment has been directly simulated with a resolved microlayer and conjugate heat transfer. The predicted bubble shapes show a good agreement with the experimental data, and the influence of dynamic contact angle models and accommodation coefficient on the long-term behavior of the bubble are discussed. These aspects were previously obscured by the challenges posed by the high computational burden. To the best of our knowledge, the present work represents the first such effort reported in the open literature. We believe the present study has effectively demonstrated the capability of the DNS solver for nucleate boiling problems. Several improvements are considered for our future work, including more advanced IHTR models and dynamic contact angle models in the context of nucleate boiling.



\backsection[Funding]{This project has received funding from the European Research Council (ERC) under the European Union’s Horizon 2020 research and innovation programme (grant agreement number 883849). We are grateful to GENCI for generous allocation on Adastra
supercomputers (grant agreement number A0152B14629) and their teams for assistance and the use of Irene-Rome at TGCC.}

\backsection[Declaration of interests]{The authors report no conflict of interest.}


\backsection[Author ORCIDs]{T. Long, https://orcid.org/0009-0006-2004-255X; J. Pan, https://orcid.org/0009-0000-5367-375X; E. Cipriano, https://orcid.org/0000-0003-4976-2578; M. Bucci, https://orcid.org/0000-0002-6423-1356; S. Zaleski, https://orcid.org/0000-0003-2004-9090.}


\appendix

\section{Ghost fluid method vs. one-fluid method}\label{appA:ghost_fluid_method_compare}
\begin{figure}
    \centerline{\includegraphics[width=0.8\textwidth]{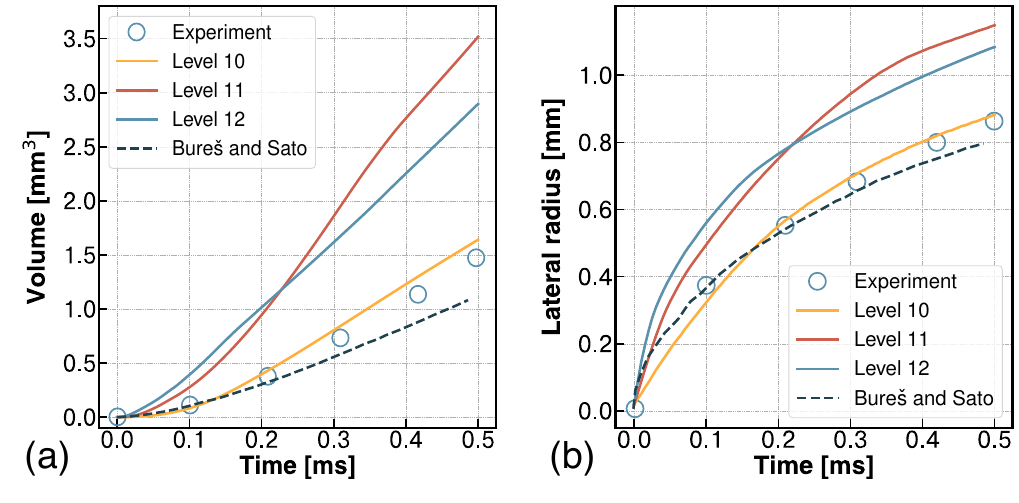}}
    \caption{The time histories of (a) the bubble volume and (b) the bubble lateral radius for case A ($\omega$ = 0.0345) at different grid levels. The results are obtained with the one-fluid method and are compared with the experimental data of \citet{bucci2020thesis} and the numerical results of \citet{burevs2022comprehensive}.}
    \label{Fig_vol_radius_a_onefluid}
\end{figure}
\begin{figure}
    \centerline{\includegraphics[width=1.0\textwidth]{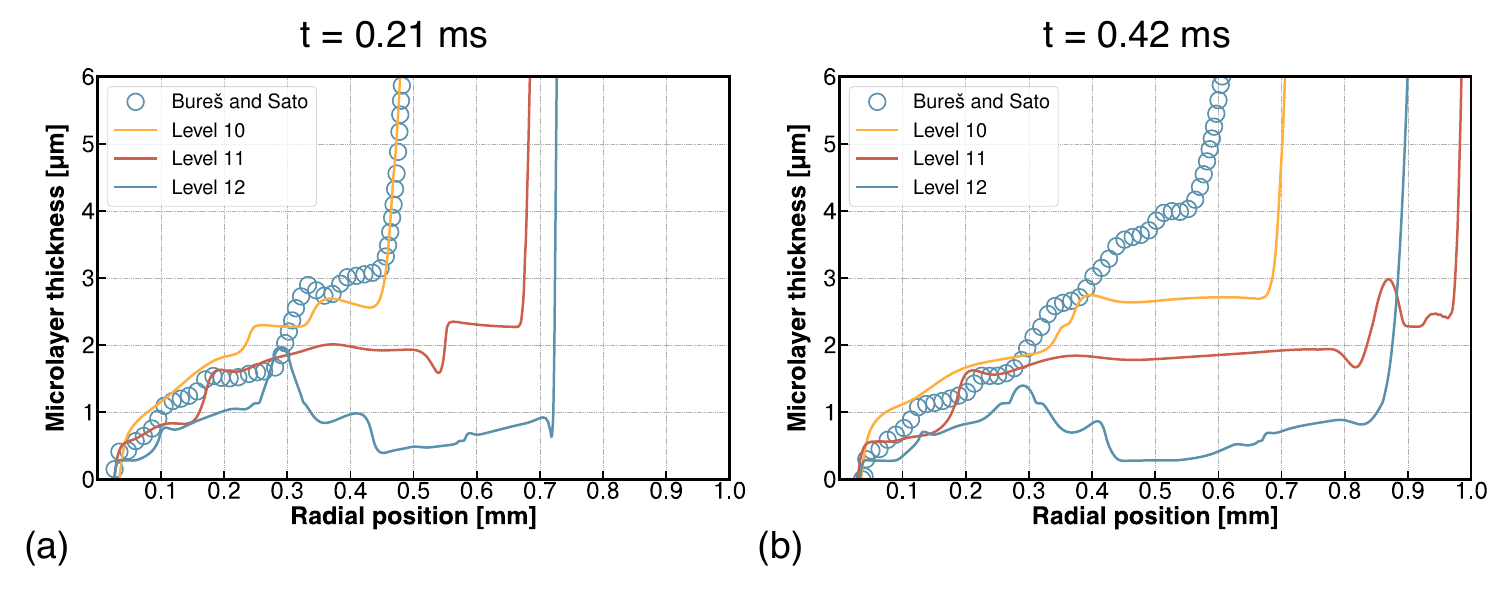}}
    \caption{The microlayer profiles at different time instants and grid levels for case A ($\omega = 0.0345$). The results are obtained with the one-fluid method and are compared with the numerical results of \citet{burevs2022comprehensive}.}
    \label{Fig_microlayer_a_onefluid}
\end{figure}
\begin{figure}
    \centerline{\includegraphics[width=0.85\textwidth]{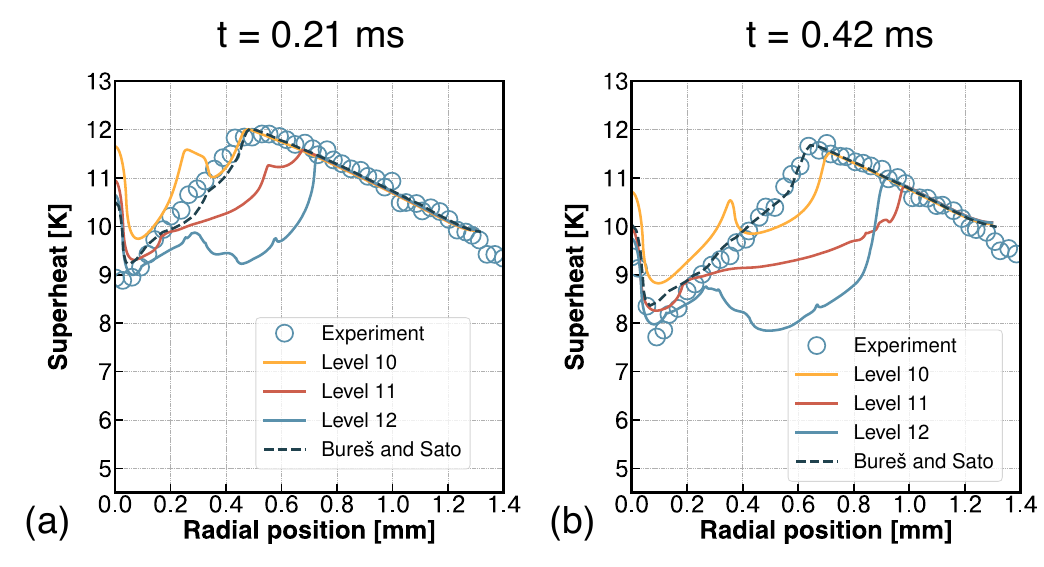}}
    \caption{The surface superheat distribution at different time instants and grid levels for case A ($\omega = 0.0345$). The results are obtained with the one-fluid method and are compared with the experimental data of \citet{bucci2020thesis} and the numerical results of \citet{burevs2022comprehensive}.}
    \label{Fig_superheat_a_onefluid}
\end{figure}
\begin{figure}
    \centerline{\includegraphics[width=0.85\textwidth]{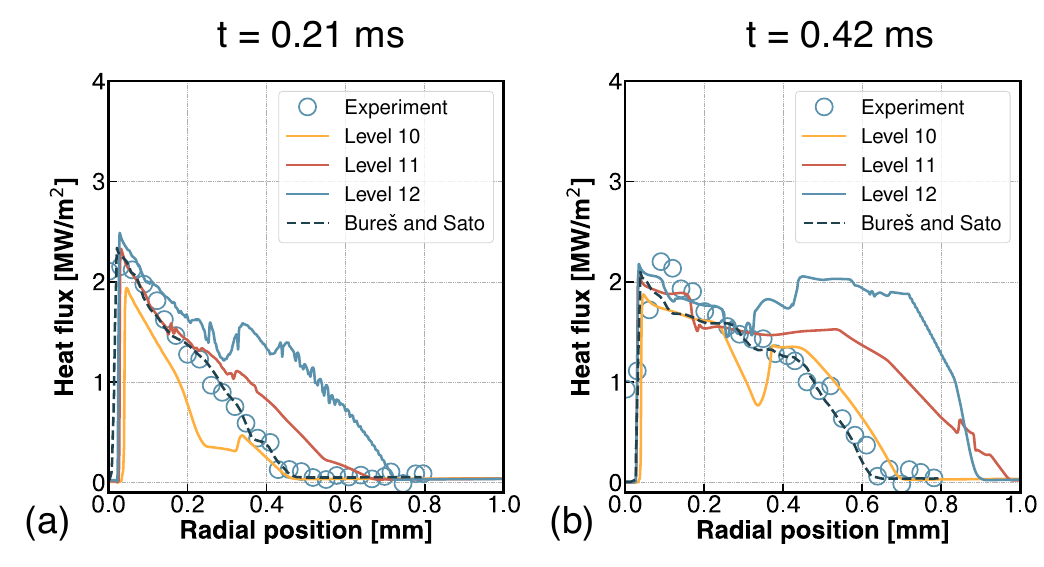}}
    \caption{The surface heat flux distribution at different time instants and grid levels for case A ($\omega = 0.0345$). The results are obtained with the one-fluid method and are compared with the experimental data of \citet{bucci2020thesis} and the numerical results of \citet{burevs2022comprehensive}.}
    \label{Fig_heatflux_a_onefluid}
\end{figure}
\begin{figure}
    \centerline{\includegraphics[width=0.75\textwidth]{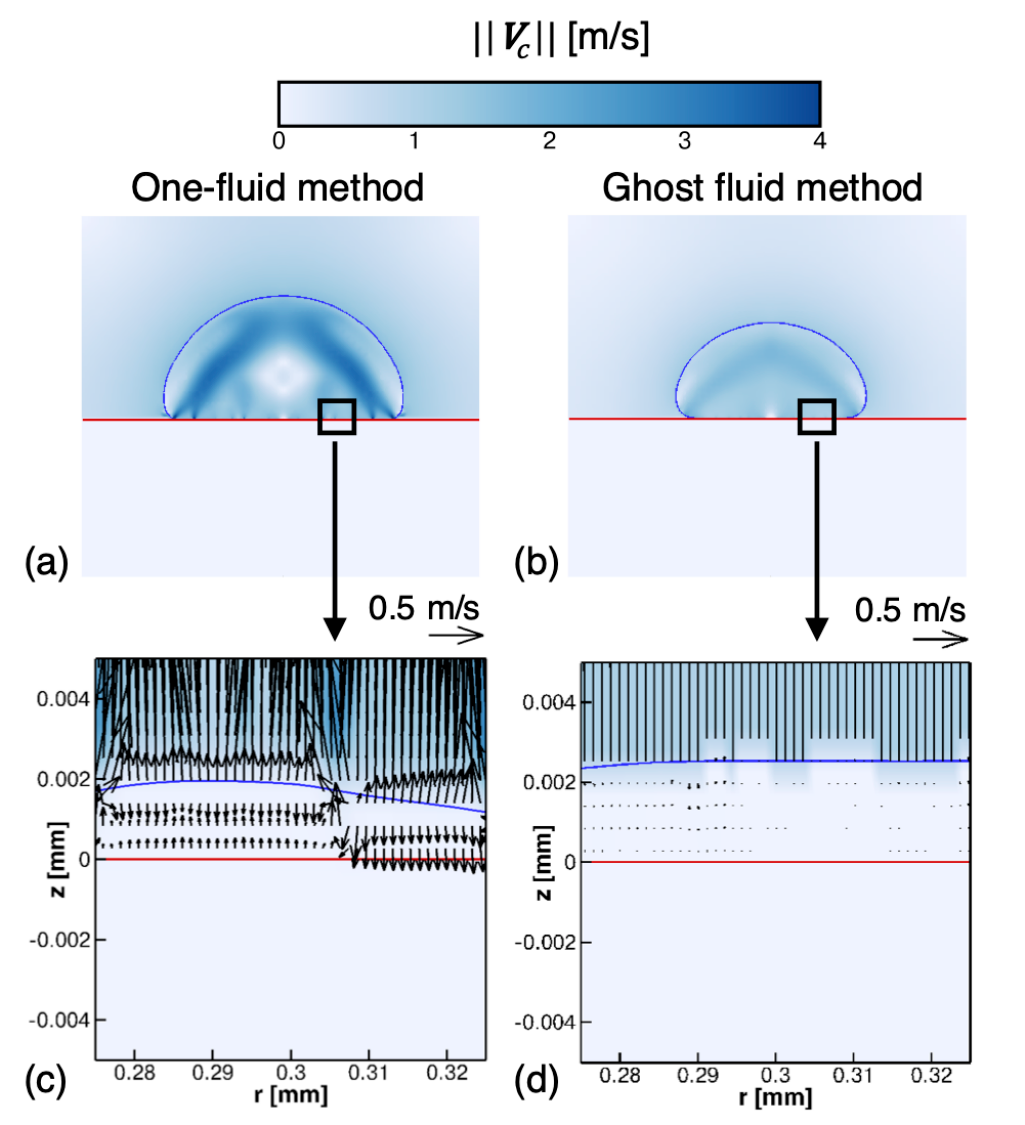}}
    \caption{The magnitudes of the cell-centered velocity $\boldsymbol{u}_c$ obtained with (a) the one-fluid method and (b) the ghost fluid method at $t = 0.21\ \rm{ms}$ and level 12. The blue line represents the liquid-vapor interface, while the red line represents the fluid-solid boundary. (c) and (d) provide close-ups of the selected region, with the velocity vectors plotted.}
    \label{Fig_vel_and_zoom}
\end{figure}
In the present study, it is evident that the computational cost can be significantly reduced using AMR. However, it is emphasized that the use of AMR necessitates a compromise: the cell-centered velocity field can only be approximately projected, leading to unphysical oscillations in the presence of phase change. Based on the well-balanced VOF framework \citep{popinet2009accurate}, the original model of \citet{cipriano2024multicomponent} performs well in a number of benchmark tests, even with the approximate projection method. However, as shown here, the original model fails when applied to nucleate boiling with a microlayer, mainly due to the intense heat and mass transfer within this thin layer. Here, the one-fluid method is compared with the ghost fluid method. Using the same setup as given in section \ref{sec4.1:simulation_setup}, case A ($\omega = 0.0345$) is simulated with the one-fluid method across grid levels 10 to 12. The time histories of the volume and lateral radius of the bubble at these grid levels are presented in figure \ref{Fig_vol_radius_a_onefluid}. It is shown that the results at grid level 10 obtained with the one-fluid method agree better with the experimental data of \citet{bucci2020thesis}. However, this better agreement is merely a numerical artifact, as increasing grid resolution leads to significant deviations rather than grid convergence. The microlayer profiles at various time instants and grid levels are depicted in figure \ref{Fig_microlayer_a_onefluid}. It can be seen that the microlayer obtained with the one-fluid method exhibits more oscillations and is thinner than that obtained with the ghost fluid method (figure \ref{Fig_microlayer_thickness}). As indicated in figures \ref{Fig_superheat_a_onefluid} and \ref{Fig_heatflux_a_onefluid}, compared to the experimental data, larger heat fluxes and smaller surface temperatures are obtained using the one-fluid method, attributed to the reduced microlayer thickness. Moreover, these oscillations in the microlayer also lead to erroneous surface temperature and heat flux distributions. 

For a better comparison between the one-fluid method and the ghost fluid method, the magnitudes of the cell-centered velocity $\boldsymbol{u}_c$ obtained with the two methods at $t = 0.21\ \rm{ms}$ and grid level 12 are given in figure \ref{Fig_vel_and_zoom}. It can be observed from figures \ref{Fig_vel_and_zoom}(a) and (b) that more pronounced numerical oscillations are introduced by the one-fluid method. When examining a selected region and comparing the velocity fields in figures \ref{Fig_vel_and_zoom}(c) and (d), we observe that the fluid within the microlayer is almost stagnant when using the ghost fluid method, consistent with findings in previous studies \citep{urbano2018direct, burevs2022comprehensive}. In contrast, significant oscillations are observed with the one-fluid method, as indicated by the erroneous vertical component of the velocity. This vertical component consequently results in an incorrect microlayer thickness.

\bibliographystyle{jfm}
\bibliography{jfm}

\begin{thebibliography}{56}
\expandafter\ifx\csname natexlab\endcsname\relax\def\natexlab#1{#1}\fi
\def\au#1{#1} \def\ed#1{#1} \def\yr#1{#1}\def\at#1{#1}\def\jt#1{\textit{#1}}
  \def\bt#1{#1}\def\bvol#1{\textbf{#1}} \def\vol#1{#1} \def\pg#1{#1}
  \def\publ#1{#1}\def\arxiv#1{#1}\def\org#1{#1}\def\st#1{\textit{#1}}

\bibitem[Afkhami \& Bussmann(2008)]{afkhami2008height}
{\sc \au{Afkhami, S.} \& \au{Bussmann, M.}} \yr{2008}  \at{Height functions for
  applying contact angles to {2D} {VOF} simulations}.  \jt{International
  Journal for Numerical Methods in Fluids}  \bvol{57}~(4),  \pg{453--472}.

\bibitem[Bell {\em et~al.\/}(1989)Bell, Colella \& Glaz]{bell1989second}
{\sc \au{Bell, J.B.}, \au{Colella, P.} \& \au{Glaz, H.M.}} \yr{1989}  \at{A
  second-order projection method for the incompressible {N}avier-{S}tokes
  equations}.  \jt{Journal of Computational Physics}  \bvol{85}~(2),
  \pg{257--283}.

\bibitem[Bucci(2020)]{bucci2020thesis}
{\sc \au{Bucci, M.}} \yr{2020}  \at{\rm{A theoretical and experimental study of
  vapor bubble dynamics in separate effect pool boiling conditions}}.
  \textit{Master's thesis}, University of Pisa, Italy, conducted at MIT.

\bibitem[Bucci {\em et~al.\/}(2016)Bucci, Richenderfer, Su, McKrell \&
  Buongiorno]{bucci2016mechanistic}
{\sc \au{Bucci, M.}, \au{Richenderfer, A.}, \au{Su, G.}, \au{McKrell, T.} \&
  \au{Buongiorno, J.}} \yr{2016}  \at{A mechanistic ir calibration technique
  for boiling heat transfer investigations}.  \jt{International Journal of
  Multiphase Flow}  \bvol{83},  \pg{115--127}.

\bibitem[Bure{\v{s}} {\em et~al.\/}(2024)Bure{\v{s}}, Bucci, Sato \&
  Bucci]{bures2024coarse}
{\sc \au{Bure{\v{s}}, L.}, \au{Bucci, M.}, \au{Sato, Y.} \& \au{Bucci, M.}}
  \yr{2024}  \at{A coarse grid approach for single bubble boiling simulations
  with the volume of fluid method}.  \jt{Computers \& Fluids}  \bvol{271},
  \pg{106182}.

\bibitem[Bure{\v{s}} \& Sato(2021)]{burevs2021modelling}
{\sc \au{Bure{\v{s}}, L.} \& \au{Sato, Y.}} \yr{2021}  \at{On the modelling of
  the transition between contact-line and microlayer evaporation regimes in
  nucleate boiling}.  \jt{Journal of Fluid Mechanics}  \bvol{916},  \pg{A53}.

\bibitem[Bure{\v{s}} \& Sato(2022)]{burevs2022comprehensive}
{\sc \au{Bure{\v{s}}, L.} \& \au{Sato, Y.}} \yr{2022}  \at{Comprehensive
  simulations of boiling with a resolved microlayer: validation and sensitivity
  study}.  \jt{Journal of Fluid Mechanics}  \bvol{933},  \pg{A54}.

\bibitem[Bur\v{e}s(2021)]{bures2021thesis}
{\sc \au{Bur\v{e}s, L.}} \yr{2021}  \at{Fundamental study on microlayer
  dynamics in nucleate boiling}. PhD thesis, L'École Polytechnique Fédérale
  de Lausanne (EPFL), Switzerland.

\bibitem[Cai {\em et~al.\/}(2024)Cai, Wang, Jiang \& Tan]{cai2024new}
{\sc \au{Cai, D.}, \au{Wang, P.}, \au{Jiang, W.} \& \au{Tan, R.}} \yr{2024}
  \at{A new microlayer depletion model for numerical simulation of bubble
  growth during nucleate boiling}.  \jt{International Journal of Heat and Mass
  Transfer}  \bvol{224},  \pg{125318}.

\bibitem[Chen {\em et~al.\/}(2023)Chen, Jin, Yu, Ling, Sun, Zhang, Jiao \&
  Tao]{chen2023modeling}
{\sc \au{Chen, Y.}, \au{Jin, S.}, \au{Yu, B.}, \au{Ling, K.}, \au{Sun, D.},
  \au{Zhang, W.}, \au{Jiao, K.} \& \au{Tao, W.}} \yr{2023}  \at{Modeling and
  study of microlayer effects on flow boiling in a mini-channel}.
  \jt{International Journal of Heat and Mass Transfer}  \bvol{208},
  \pg{124039}.

\bibitem[Chen {\em et~al.\/}(2024)Chen, Yu, Lu, Wang, Sun, Jiao, Zhang \&
  Tao]{chen2024review}
{\sc \au{Chen, Y.}, \au{Yu, B.}, \au{Lu, W.}, \au{Wang, B.}, \au{Sun, D.},
  \au{Jiao, K.}, \au{Zhang, W.} \& \au{Tao, W.}} \yr{2024}  \at{Review on
  numerical simulation of boiling heat transfer from atomistic to mesoscopic
  and macroscopic scales}.  \jt{International Journal of Heat and Mass
  Transfer}  \bvol{225},  \pg{125396}.

\bibitem[Cheng \& Xia(2017)]{cheng2017fundamental}
{\sc \au{Cheng, L.} \& \au{Xia, G.}} \yr{2017}  \at{Fundamental issues,
  mechanisms and models of flow boiling heat transfer in microscale channels}.
  \jt{International Journal of Heat and Mass Transfer}  \bvol{108},
  \pg{97--127}.

\bibitem[Cipriano(2023)]{edosandbox}
{\sc \au{Cipriano, E.}} \yr{2023} Code repository.
  \url{http://basilisk.fr/sandbox/ecipriano/}.

\bibitem[Cipriano {\em et~al.\/}(2024)Cipriano, Frassoldati, Faravelli, Popinet
  \& Cuoci]{cipriano2024multicomponent}
{\sc \au{Cipriano, E.}, \au{Frassoldati, A.}, \au{Faravelli, T.}, \au{Popinet,
  S.} \& \au{Cuoci, A.}} \yr{2024}  \at{Multicomponent droplet evaporation in a
  geometric volume-of-fluid framework}.  \jt{Journal of Computational Physics}
  \bvol{507},  \pg{112955}.

\bibitem[Dhruv {\em et~al.\/}(2019)Dhruv, Balaras, Riaz \&
  Kim]{dhruv2019formulation}
{\sc \au{Dhruv, A.}, \au{Balaras, E.}, \au{Riaz, A.} \& \au{Kim, J.}} \yr{2019}
   \at{A formulation for high-fidelity simulations of pool boiling in low
  gravity}.  \jt{International Journal of Multiphase Flow}  \bvol{120},
  \pg{103099}.

\bibitem[Ding {\em et~al.\/}(2018)Ding, Krepper \& Hampel]{ding2018evaluation}
{\sc \au{Ding, W.}, \au{Krepper, E.} \& \au{Hampel, U.}} \yr{2018}
  \at{Evaluation of the microlayer contribution to bubble growth in horizontal
  pool boiling with a mechanistic model that considers dynamic contact angle
  and base expansion}.  \jt{International Journal of Heat and Fluid Flow}
  \bvol{72},  \pg{274--287}.

\bibitem[El~Mellas {\em et~al.\/}(2024)El~Mellas, Samkhaniani, Falsetti, Stroh,
  Icardi \& Magnini]{el2024numerical}
{\sc \au{El~Mellas, I.}, \au{Samkhaniani, N.}, \au{Falsetti, C.}, \au{Stroh,
  A.}, \au{Icardi, M.} \& \au{Magnini, M.}} \yr{2024}  \at{Numerical
  investigation of bubble dynamics and flow boiling heat transfer in
  cylindrical micro-pin-fin heat exchangers}.  \jt{International Journal of
  Heat and Mass Transfer}  \bvol{228},  \pg{125620}.

\bibitem[Fang {\em et~al.\/}(2008)Fang, Hidrovo, Wang, Eaton \&
  Goodson]{fang20083Dcontactline}
{\sc \au{Fang, C.}, \au{Hidrovo, C.}, \au{Wang, F.}, \au{Eaton, J.} \&
  \au{Goodson, K.}} \yr{2008}  \at{3-{D} numerical simulation of contact angle
  hysteresis for microscale two phase flow}.  \jt{International Journal of
  Multiphase Flow}  \bvol{34}~(7),  \pg{690--705}.

\bibitem[Giustini(2024)]{giustini2024hydrodynamic}
{\sc \au{Giustini, G.}} \yr{2024}  \at{Hydrodynamic analysis of liquid
  microlayer formation in nucleate boiling of water}.  \jt{International
  Journal of Multiphase Flow}  \bvol{172},  \pg{104718}.

\bibitem[Giustini {\em et~al.\/}(2016)Giustini, Jung, Kim \&
  Walker]{giustini2016evaporative}
{\sc \au{Giustini, G.}, \au{Jung, S.}, \au{Kim, H.} \& \au{Walker, S.P.}}
  \yr{2016}  \at{Evaporative thermal resistance and its influence on
  microscopic bubble growth}.  \jt{International Journal of Heat and Mass
  Transfer}  \bvol{101},  \pg{733--741}.

\bibitem[Giustini {\em et~al.\/}(2020)Giustini, Kim, Issa \&
  Bluck]{giustini2020modelling}
{\sc \au{Giustini, G.}, \au{Kim, H.}, \au{Issa, R.I.} \& \au{Bluck, M.J.}}
  \yr{2020}  \at{Modelling microlayer formation in boiling sodium}.
  \jt{Fluids}  \bvol{5}~(4),  \pg{213}.

\bibitem[Guion {\em et~al.\/}(2018)Guion, Afkhami, Zaleski \&
  Buongiorno]{guion2018simulations}
{\sc \au{Guion, A.}, \au{Afkhami, S.}, \au{Zaleski, S.} \& \au{Buongiorno, J.}}
  \yr{2018}  \at{Simulations of microlayer formation in nucleate boiling}.
  \jt{International Journal of Heat and Mass Transfer}  \bvol{127},
  \pg{1271--1284}.

\bibitem[H{\"a}nsch \& Walker(2016)]{hansch2016hydrodynamics}
{\sc \au{H{\"a}nsch, S.} \& \au{Walker, S.}} \yr{2016}  \at{The hydrodynamics
  of microlayer formation beneath vapour bubbles}.  \jt{International Journal
  of Heat and Mass Transfer}  \bvol{102},  \pg{1282--1292}.

\bibitem[H{\"a}nsch \& Walker(2019)]{hansch2019microlayer}
{\sc \au{H{\"a}nsch, S.} \& \au{Walker, S.}} \yr{2019}  \at{Microlayer
  formation and depletion beneath growing steam bubbles}.  \jt{International
  Journal of Multiphase Flow}  \bvol{111},  \pg{241--263}.

\bibitem[Huber {\em et~al.\/}(2017)Huber, Tanguy, Sagan \&
  Colin]{huber2017direct}
{\sc \au{Huber, G.}, \au{Tanguy, S.}, \au{Sagan, M.} \& \au{Colin, C.}}
  \yr{2017}  \at{Direct numerical simulation of nucleate pool boiling at large
  microscopic contact angle and moderate {Jakob} number}.  \jt{International
  Journal of Heat and Mass Transfer}  \bvol{113},  \pg{662--682}.

\bibitem[Jung \& Kim(2014)]{jung2014experimental}
{\sc \au{Jung, S.} \& \au{Kim, H.}} \yr{2014}  \at{An experimental method to
  simultaneously measure the dynamics and heat transfer associated with a
  single bubble during nucleate boiling on a horizontal surface}.
  \jt{International Journal of Heat and Mass Transfer}  \bvol{73},
  \pg{365--375}.

\bibitem[Jung \& Kim(2018)]{jung2018hydrodynamic}
{\sc \au{Jung, S.} \& \au{Kim, H.}} \yr{2018}  \at{Hydrodynamic formation of a
  microlayer underneath a boiling bubble}.  \jt{International Journal of Heat
  and Mass Transfer}  \bvol{120},  \pg{1229--1240}.

\bibitem[Kim {\em et~al.\/}(2020)Kim, Sergis, Kim \&
  Hardalupas]{kim2020assessing}
{\sc \au{Kim, M.}, \au{Sergis, A.}, \au{Kim, S.} \& \au{Hardalupas, Y.}}
  \yr{2020}  \at{Assessing the accuracy of the heat flux measurement for the
  study of boiling phenomena}.  \jt{International Journal of Heat and Mass
  Transfer}  \bvol{148},  \pg{119019}.

\bibitem[Long(2024)]{tiansandbox}
{\sc \au{Long, T.}} \yr{2024} Code repository.
  \url{http://basilisk.fr/sandbox/tianlong/}.

\bibitem[Long {\em et~al.\/}(2024)Long, Pan \& Zaleski]{long2024edge}
{\sc \au{Long, T.}, \au{Pan, J.} \& \au{Zaleski, S.}} \yr{2024}  \at{An
  {E}dge-based {I}nterface {T}racking ({EBIT}) method for multiphase flows with
  phase change}.  \jt{Journal of Computational Physics}  \bvol{513},
  \pg{113159}.

\bibitem[Manglik(2006)]{manglik2006advancements}
{\sc \au{Manglik, R.M.}} \yr{2006}  \at{On the advancements in boiling,
  two-phase flow heat transfer, and interfacial phenomena}.  \jt{Journal of
  Heat Transfer}  \bvol{128}~(12),  \pg{1237--1242}.

\bibitem[Marek \& Straub(2001)]{marek2001analysis}
{\sc \au{Marek, R.} \& \au{Straub, J.}} \yr{2001}  \at{Analysis of the
  evaporation coefficient and the condensation coefficient of water}.
  \jt{International Journal of Heat and Mass Transfer}  \bvol{44}~(1),
  \pg{39--53}.

\bibitem[Narayan \& Srivastava(2021)]{narayan2021non}
{\sc \au{Narayan, L.S.} \& \au{Srivastava, A.}} \yr{2021}  \at{Non-contact
  experiments to quantify the microlayer evaporation heat transfer coefficient
  during isolated nucleate boiling regime}.  \jt{International Communications
  in Heat and Mass Transfer}  \bvol{122},  \pg{105191}.

\bibitem[Nathanson {\em et~al.\/}(1996)Nathanson, Davidovits, Worsnop \&
  Kolb]{nathanson1996dynamics}
{\sc \au{Nathanson, G.M.}, \au{Davidovits, P.}, \au{Worsnop, D.R.} \& \au{Kolb,
  C.E.}} \yr{1996}  \at{Dynamics and kinetics at the gas- liquid interface}.
  \jt{The Journal of Physical Chemistry}  \bvol{100}~(31),  \pg{13007--13020}.

\bibitem[Pan {\em et~al.\/}(2024)Pan, Long, Chirco, Scardovelli, Popinet \&
  Zaleski]{pan2023edge}
{\sc \au{Pan, J.}, \au{Long, T.}, \au{Chirco, L.}, \au{Scardovelli, R.},
  \au{Popinet, S.} \& \au{Zaleski, S.}} \yr{2024}  \at{An edge-based interface
  tracking ({EBIT}) method for multiphase-flows simulation with surface
  tension}.  \jt{Journal of Computational Physics}  \bvol{508},  \pg{113016}.

\bibitem[Persad \& Ward(2016)]{persad2016expressions}
{\sc \au{Persad, A.H.} \& \au{Ward, C.A.}} \yr{2016}  \at{Expressions for the
  evaporation and condensation coefficients in the {H}ertz-{K}nudsen relation}.
   \jt{Chemical Reviews}  \bvol{116}~(14),  \pg{7727--7767}.

\bibitem[Popinet(2003)]{popinet2003gerris}
{\sc \au{Popinet, S.}} \yr{2003}  \at{Gerris: a tree-based adaptive solver for
  the incompressible euler equations in complex geometries}.  \jt{Journal of
  Computational Physics}  \bvol{190}~(2),  \pg{572--600}.

\bibitem[Popinet(2009)]{popinet2009accurate}
{\sc \au{Popinet, S.}} \yr{2009}  \at{An accurate adaptive solver for
  surface-tension-driven interfacial flows}.  \jt{Journal of Computational
  Physics}  \bvol{228}~(16),  \pg{5838--5866}.

\bibitem[Popinet(2015)]{popinet2015quadtree}
{\sc \au{Popinet, S.}} \yr{2015}  \at{A quadtree-adaptive multigrid solver for
  the {S}erre--{G}reen--{N}aghdi equations}.  \jt{Journal of Computational
  Physics}  \bvol{302},  \pg{336--358}.

\bibitem[Roache(1998)]{roache1998verification}
{\sc \au{Roache, P.J.}} \yr{1998} {\em Verification and validation in
  computational science and engineering\/}, ,  \vol{vol. 895}.  \publ{Hermosa
  Albuquerque, NM}.

\bibitem[Saini {\em et~al.\/}(2024)Saini, Chen, Zaleski \&
  Fuster]{saini2024direct}
{\sc \au{Saini, M.}, \au{Chen, X.}, \au{Zaleski, S.} \& \au{Fuster, D.}}
  \yr{2024}  \at{Direct numerical simulations of microlayer formation during
  heterogeneous bubble nucleation}.  \jt{Journal of Fluid Mechanics}
  \bvol{984},  \pg{A70}.

\bibitem[Sato \& Niceno(2015)]{sato2015depletable}
{\sc \au{Sato, Y.} \& \au{Niceno, B.}} \yr{2015}  \at{A depletable micro-layer
  model for nucleate pool boiling}.  \jt{Journal of Computational physics}
  \bvol{300},  \pg{20--52}.

\bibitem[Scriven(1959)]{scriven1959dynamics}
{\sc \au{Scriven, L.E.}} \yr{1959}  \at{On the dynamics of phase growth}.
  \jt{Chemical Engineering Science}  \bvol{10}~(1–2),  \pg{1--13}.

\bibitem[Tanguy {\em et~al.\/}(2014)Tanguy, Sagan, Lalanne, Couderc \&
  Colin]{tanguy2014benchmarks}
{\sc \au{Tanguy, S.}, \au{Sagan, M.}, \au{Lalanne, B.}, \au{Couderc, F.} \&
  \au{Colin, C.}} \yr{2014}  \at{Benchmarks and numerical methods for the
  simulation of boiling flows}.  \jt{Journal of Computational Physics}
  \bvol{264},  \pg{1--22}.

\bibitem[Tecchio {\em et~al.\/}(2024)Tecchio, Zhang, Cariteau, Zalczer, Bulkin,
  Charliac, Vassant \& Nikolayev]{tecchio2024microlayer}
{\sc \au{Tecchio, C.}, \au{Zhang, X.}, \au{Cariteau, B.}, \au{Zalczer, G.},
  \au{Bulkin, P.}, \au{Charliac, J.}, \au{Vassant, S.} \& \au{Nikolayev, V.S.}}
  \yr{2024}  \at{Microlayer in nucleate boiling seen as {L}andau-{L}evich film
  with dewetting and evaporation}.  \jt{Journal of Fluid Mechanics}
  \bvol{989},  \pg{A4}.

\bibitem[Torres {\em et~al.\/}(2024)Torres, Urbano, Colin \&
  Tanguy]{torres2024coupling}
{\sc \au{Torres, L.}, \au{Urbano, A.}, \au{Colin, C.} \& \au{Tanguy, S.}}
  \yr{2024}  \at{On the coupling between direct numerical simulation of
  nucleate boiling and a micro-region model at the contact line}.  \jt{Journal
  of Computational Physics}  \bvol{497},  \pg{112602}.

\bibitem[Urbano {\em et~al.\/}(2018)Urbano, Tanguy, Huber \&
  Colin]{urbano2018direct}
{\sc \au{Urbano, A.}, \au{Tanguy, S.}, \au{Huber, G.} \& \au{Colin, C.}}
  \yr{2018}  \at{Direct numerical simulation of nucleate boiling in micro-layer
  regime}.  \jt{International Journal of Heat and Mass Transfer}  \bvol{123},
  \pg{1128--1137}.

\bibitem[Utaka {\em et~al.\/}(2013)Utaka, Kashiwabara \&
  Ozaki]{utaka2013microlayer}
{\sc \au{Utaka, Y.}, \au{Kashiwabara, Y.} \& \au{Ozaki, M.}} \yr{2013}
  \at{Microlayer structure in nucleate boiling of water and ethanol at
  atmospheric pressure}.  \jt{International Journal of Heat and Mass Transfer}
  \bvol{57}~(1),  \pg{222--230}.

\bibitem[Wang {\em et~al.\/}(2023)Wang, Liu, Bayeul-Lain{\'e}, Murphy, Katz \&
  Coutier-Delgosha]{wang2023analysis}
{\sc \au{Wang, H.}, \au{Liu, S.}, \au{Bayeul-Lain{\'e}, A.C.}, \au{Murphy, D.},
  \au{Katz, J.} \& \au{Coutier-Delgosha, O.}} \yr{2023}  \at{Analysis of
  high-speed drop impact onto deep liquid pool}.  \jt{Journal of Fluid
  Mechanics}  \bvol{972},  \pg{A31}.

\bibitem[Weymouth \& Yue(2010)]{weymouth2010conservative}
{\sc \au{Weymouth, G.D.} \& \au{Yue, D.K.P.}} \yr{2010}  \at{Conservative
  volume-of-fluid method for free-surface simulations on cartesian-grids}.
  \jt{Journal of Computational Physics}  \bvol{229}~(8),  \pg{2853--2865}.

\bibitem[Yabuki \& Nakabeppu(2014)]{yabuki2014heat}
{\sc \au{Yabuki, T.} \& \au{Nakabeppu, O.}} \yr{2014}  \at{Heat transfer
  mechanisms in isolated bubble boiling of water observed with mems sensor}.
  \jt{International Journal of Heat and Mass Transfer}  \bvol{76},
  \pg{286--297}.

\bibitem[Zhang {\em et~al.\/}(2023{\natexlab{{\em a\/}}})Zhang, Rafique, Ding,
  Bolotnov \& Hampel]{zhang2023direct}
{\sc \au{Zhang, J.}, \au{Rafique, M.}, \au{Ding, W.}, \au{Bolotnov, I.A.} \&
  \au{Hampel, U.}} \yr{2023{\natexlab{{\em a\/}}}}  \at{Direct numerical
  simulation of microlayer formation and evaporation underneath a growing
  bubble driven by the local temperature gradient in nucleate boiling}.
  \jt{International Journal of Thermal Sciences}  \bvol{193},  \pg{108551}.

\bibitem[Zhang {\em et~al.\/}(2023{\natexlab{{\em b\/}}})Zhang, Rafique, Ding,
  Bolotnov \& Hampel]{zhang2023directpillar}
{\sc \au{Zhang, J.}, \au{Rafique, M.}, \au{Ding, W.}, \au{Bolotnov, I.A.} \&
  \au{Hampel, U.}} \yr{2023{\natexlab{{\em b\/}}}}  \at{A direct numerical
  simulation study to elucidate the enhancement of heat transfer for nucleate
  boiling on surfaces with micro-pillars}.  \jt{International Communications in
  Heat and Mass Transfer}  \bvol{147},  \pg{106943}.

\bibitem[Zhao {\em et~al.\/}(2022)Zhao, Zhang \& Ni]{zhao2022boiling}
{\sc \au{Zhao, S.}, \au{Zhang, J.} \& \au{Ni, M.J.}} \yr{2022}  \at{Boiling and
  evaporation model for liquid-gas flows: A sharp and conservative method based
  on the geometrical vof approach}.  \jt{Journal of Computational Physics}
  \bvol{452},  \pg{110908}.

\bibitem[Zhao {\em et~al.\/}(2002)Zhao, Masuoka \& Tsuruta]{zhao2002unified}
{\sc \au{Zhao, Y.}, \au{Masuoka, T.} \& \au{Tsuruta, T.}} \yr{2002}
  \at{Unified theoretical prediction of fully developed nucleate boiling and
  critical heat flux based on a dynamic microlayer model}.  \jt{International
  Journal of Heat and Mass Transfer}  \bvol{45}~(15),  \pg{3189--3197}.

\bibitem[Zupan{\v{c}}i{\v{c}} {\em et~al.\/}(2022)Zupan{\v{c}}i{\v{c}},
  Gregor{\v{c}}i{\v{c}}, Bucci, Wang, Aguiar \& Bucci]{zupanvcivc2022wall}
{\sc \au{Zupan{\v{c}}i{\v{c}}, M.}, \au{Gregor{\v{c}}i{\v{c}}, P.}, \au{Bucci,
  M.}, \au{Wang, C.}, \au{Aguiar, G.M.} \& \au{Bucci, M.}} \yr{2022}  \at{The
  wall heat flux partitioning during the pool boiling of water on thin metallic
  foils}.  \jt{Applied Thermal Engineering}  \bvol{200},  \pg{117638}.

\end{thebibliography}

\end{document}